\documentclass[lettersize,journal]{IEEEtran}
\usepackage{amsmath,amsfonts}
\usepackage{array}
\usepackage{textcomp}
\usepackage{stfloats}
\usepackage{url}
\usepackage{verbatim}
\usepackage{graphicx}
\usepackage{cite}
\usepackage{tikz}
\usepackage{makecell}
\usepackage{enumitem}
\usepackage{subfig}
\usepackage{pifont}
\usepackage[linesnumbered,ruled,vlined]{algorithm2e}
\newcommand{\cmark}{\ding{51}}  
\newcommand{\xmark}{\ding{55}}  
\newcommand{\pmark}{$\sim$}   

\begin{document}

\title{Cyber Attacks Detection, Prevention, and Source Localization in Digital Substation Communication using Hybrid Statistical-Deep Learning}

\author{Nicola Cibin, Bas Mulder, Herman Carstens, Junho Hong, Alexandru Ştefanov 
\thanks{This work was supported in part by the EU Horizon 2020 Marie Sklodowska Curie InnoCyPES Project with Grant Agreement No. 956433, and in part by the EU Horizon Europe ESTELAR Project with Grant Agreement No. 101192574. (Corresponding author: Alexandru Ştefanov)}
\thanks{Nicola Cibin and Alexandru Ştefanov are with the Department of Electrical Sustainable Energy (ESE), Delft University of Technology, Mekelweg 4, 2628CD, Delft, The Netherlands (e-mails: \{n.cibin, a.i.stefanov\}@tudelft.nl).}
\thanks{Bas Mulder is with DNV Netherlands B.V., Utrechtseweg 310, 6812AR Arnhem, The Netherlands (e-mail: bas.mulder@dnv.com).}
\thanks{Herman Carstens is with Elia Group Innovation, Keizerslaan 20, 1000 Brussel, Belgium (e-mail: herman.carstens@elia.be).}
\thanks{Junho Hong is with the Department of Electrical and Computer Engineering, University of Michigan–Dearborn, Dearborn, MI 48128, USA (e-mail: jhwr@umich.edu).}}

\maketitle

\begin{abstract}
The digital transformation of power systems is accelerating the adoption of IEC 61850 standard. However, its communication protocols lack built-in authentication and encryption, leaving them vulnerable to Man-in-the-Middle (MitM) and malicious frame injection attacks that can disrupt protection schemes operation. While most existing research focuses on detecting cyber attacks in digital substations, intrusion prevention systems have been largely overlooked due to concerns about potential network disruptions. To address this gap, this paper proposes an integrated hybrid statistical-deep learning method for detecting, preventing, and localizing IEC 61850 Sampled Values (SV)-based cyber attacks. The method models SV frames arrival times using exponentially modified Gaussian distributions and prevents malicious frames from reaching the targeted Intelligent Electronic Devices (IEDs). Malicious frames are dropped with minimal processing overhead and latency, while the method remains robust to network latency, jitter, and time-synchronization issues, and ensures a near-zero false positive rate under non-attack conditions. Long short-term memory and Elman recurrent neural networks are used to identify anomalous variations in the estimated probability distributions across IEDs for detecting and localizing MitM attacks on SV streams. The proposed method is validated across three testbeds comprising industrial-grade communication and protection devices, hardware-in-the-loop simulations, virtualized IEDs and merging units, and high-fidelity emulated networks. Results demonstrate the method’s practicality and effectiveness for deployment in IEC 61850-compliant digital substations.
\end{abstract}


\newcommand\submittedtext{%
  \footnotesize This work has been submitted to the IEEE for possible publication. Copyright may be transferred without notice, after which this version may no longer be accessible.}

\newcommand\submittednotice{%
\begin{tikzpicture}[remember picture,overlay]
\node[anchor=south,yshift=10pt] at (current page.south) {\fbox{\parbox{\dimexpr0.65\textwidth-\fboxsep-\fboxrule\relax}{\submittedtext}}};
\end{tikzpicture}%
}

\submittednotice

\section{Introduction}
As proved by the latest cyber attacks on Ukraine's power grid in 2015, 2016, and 2022 \cite{case2016ukraine, slowik2019crashoverride, kozak_industroyer_2023}, and the attempted one on the United Kingdom’s power grid in 2020 \cite{cimpanu2024middleman}, it is crucial to provide Industrial Control Systems (ICS), and, in broader terms, Operational Technology (OT) infrastructures, with state-of-the-art cyber security controls to ensure a secure and resilient power system operation. The growing integration of Information Technology (IT) and OT infrastructures enables efficient power system operation and management. However, it also expands the attack surface and increases the risk of cyber attacks by introducing new cyber security threats \cite{christopher2024sans}. 

Communication protocols defined in IEC 61850 standard have been developed to support advanced protection, automation, and control in digital substations. However, IEC 61850 inherently lacks fundamental network security mechanisms such as authentication and encryption. This deficiency is especially evident in the IEC 61850 Sampled Values (SV) protocol, where the absence of cryptographic protection allows malicious actors to inject forged frames into the substation OT communication network, potentially leading to major disruptions such as delayed fault clearance or unintended circuit breaker operations. Furthermore, these OT disruptions can affect power system stability, cause cascading failures and lead to large-scale outages or even a complete blackout \cite{rajkumar_cyber_2020, rajkumar_dynamical_2024}.
Motivated by these risks, several works in the literature have proposed the adoption of authentication and encryption to secure SV frames \cite{hussain_effective_2023, hong_implementation_2022, elrawy_geometrical_2024, rodriguez_fixed-latency_2021, suhail_hussain_lightweight_2024, presekal2026enhancing}. Cryptographic schemes provide strong guarantees of authenticity and integrity for SV frames, addressing the protocol's fundamental security deficiencies at the message level. However, they raise several challenges and limitations, including additional computational load, communication latency, message overhead, and increased infrastructure complexity. Other viable options to enhance cyber security in digital substations are the deployment of Intrusion Detection and Intrusion Prevention Systems (IDSs and IPSs, respectively).

While existing works mainly focused on IDSs \cite{gaspar_smart_2023}, the deployment of IPSs in digital substations has so far received limited attention, mainly due to a general reluctance to adopt such solutions in OT infrastructures. This reluctance primarily stems from the risk of potential disruptions to time-critical OT communication networks in digital substations, including increased latency, packet loss, and additional computational overhead. However, cyber attacks detection, prevention, and source localization address fundamentally different problems and entail distinct technical challenges. Detection identifies the presence of an attack and raises an alert, but the malicious frames still reach the targeted Intelligent Electronic Device (IED), where they may already have affected protection and control functions before any mitigation action is taken. The main technical challenge for detection is achieving high accuracy with a low False Positive Rate (FPR) to prevent overwhelming the security operation center with false alarms while providing prompt detection of threats. Prevention, in contrast, requires per-frame discrimination between legitimate and malicious OT communication traffic in real-time, so that malicious frames can be dropped before reaching the protected device. This introduces the additional and more demanding technical challenge of ensuring that the discrimination mechanism does not introduce processing delays exceeding the IEC 61850-5 and IEC 61869-9 latency requirements, and drops the minimum number of legitimate frames as possible under any operating condition. Source localization addresses an even more demanding problem, namely, identifying which specific device has been compromised and is being exploited to further target other systems in the OT communication network. This enables targeted incident investigation and mitigation actions such as isolating the compromised IED. The technical challenge here lies in inferring the position of the attacker from observable OT traffic alone, without relying on additional infrastructure such as Software-Defined Networking (SDN) \cite{girdhar2025sdn}. Therefore, while detection alone is not sufficient to ensure a resilient power system operation, proactive intrusion prevention mechanisms paired with attack source localization capabilities are essential to mitigate undesired consequences before they propagate through the OT system, while ensuring no interference with protection, automation, and control functionalities.

These challenges of intrusion detection and prevention in OT systems have motivated significant research across multiple domains. Outside the digital substation context specifically, Deep Learning (DL) has been widely applied to anomaly detection. For instance, Zeng et al. \cite{zeng2025evolutionary} proposed an evolutionary adversarial autoencoder for unsupervised anomaly detection in Industrial Internet of Things (IIoT) systems; similarly, Lu et al. \cite{lu2025moarnn} introduced a multi-objective automated Recurrent Neural Network (RNN) with attention mechanism for cyber attack detection in unmanned aerial vehicles. Within the power system domain, DL has also been applied to attack detection and localization at the state estimation level: Amirian et al.  \cite{amirian2025dynamic} proposed dynamic variational autoencoders for enhanced false data injection attacks in power system state estimation, and Lu et al. \cite{lu2024representation} proposed representation learning-based convolutional neural networks for false data injection attack localization in wide-area networks. While these methods demonstrate the value of automated and unsupervised DL architectures for cyber attack detection in OT systems, they do not address the protocol-specific timing constraints of IEC 61850-compliant digital substations, where intrusion prevention must operate within strict latency requirements and must not introduce communication network disruptions.

Focusing on IEC 61850 digital substations specifically, multiple studies have proposed intrusion detection methods based on the current and voltage measurements reported in SV frames \cite{ustun_artificial_2021, mo_sampled_2023, el_hariri_iec_2019, hussain_novel_2023, narag2024deep}. These methods leverage the underlying physical power system model, which provides a robust basis for detecting attacks that produce physically implausible measurements. Nonetheless, a key limitation of these IDSs is that relying solely on reported measurements can be misleading during replay attacks, in which normal operating conditions are re-injected by the attacker. In such cases, the measurements appear consistent with the physical model used for validation, and SV attack detection fails. To overcome this limitation, other authors have proposed IDSs that consider communication network traffic and SV frames header, relying on information such as the number of SV frames received each second ($FS$), the value of the \texttt{smpCnt} field, and the Inter-Frame Arrival Time (I-FAT) \cite{hong_intelligent_2019, delhomme2024dos, hussain_novel_2023, wannous_analysis_2019, hong_detection_2014, eynawi_machine_2024, manzoor_zero-day_2024, hong_integrated_2014}. While these detection methods do not require modeling the power system's physical behavior and can effectively identify basic SV injection attacks, they do not work against more sophisticated attacks, such as replay with spoofing and Man-in-the-Middle (MitM). In particular, when an attacker compromises a forwarding device, blocks the legitimate SV stream, and injects a malicious one, detection becomes infeasible because the attacker can simultaneously maintain the expected SV frames publishing rate ($FS$), increment the \texttt{smpCnt} field consistently, spoof the SV stream identifier, and closely match the expected inter-frame arrival timing. Moreover, the time required to detect the attacks ranges from a few hundred microseconds \cite{ustun_artificial_2021} to hundreds of milliseconds \cite{manzoor_zero-day_2024}, depending on the available computational power and the implemented detection method. For these reasons, existing approaches cannot be applied for real-time intrusion prevention, MitM attacks detection, and malicious device localization. Table \ref{tab:comparison1} summarizes the SV streams features used by the intrusion detection methods previously presented in the literature.

\begin{table}[!t]
\caption{SV frames features used for cyber attacks detection in digital substations.}
\label{tab:comparison1}
\centering
\begin{tabular}{lcccccc}
\hline
 & \textbf{PHY} & \textbf{FS} & \textbf{smpCnt} & \textbf{I-FAT} & $\mathbf{F_{as}}$\\
\hline
Ustun et al. \cite{ustun_artificial_2021} & \cmark & \xmark & \xmark & \xmark & \xmark\\
Mo et al. \cite{mo_sampled_2023} & \cmark & \xmark & \xmark & \xmark & \xmark\\
El Hariri et al. \cite{el_hariri_iec_2019} & \cmark & \xmark & \xmark & \xmark & \xmark\\
Hussain et al. \cite{hussain_novel_2023} & \cmark & \xmark & \cmark & \xmark & \xmark\\
Narag et al. \cite{narag2024deep} & \cmark & \xmark & \xmark & \xmark & \xmark\\
Hong et al. \cite{hong_intelligent_2019} & \xmark & \cmark & \cmark & \xmark & \xmark\\
Delhomme et al. \cite{delhomme2024dos} & \xmark & \xmark & \cmark & \cmark & \xmark\\
Wannous et al. \cite{wannous_analysis_2019} & \xmark & \cmark & \xmark & \cmark & \xmark\\
Hong et al. \cite{hong_integrated_2014,hong_detection_2014} & \xmark & \cmark & \cmark & \xmark & \xmark\\
Eynawi et al. \cite{eynawi_machine_2024} & \cmark & \xmark & \xmark & \cmark & \xmark\\
Manzoor et al. \cite{manzoor_zero-day_2024} & \cmark & \xmark & \cmark & \xmark & \xmark\\
\hline
\textbf{Our Solution} & \textbf{\xmark} & \textbf{\cmark} & \textbf{\cmark} & \textbf{\cmark} & \textbf{\cmark}\\
\hline
\end{tabular}
\end{table}

Among these limitations, the challenge of attack source localization in digital substations remains largely unexplored. To the best of the authors’ knowledge, only three prior works have addressed this issue by leveraging SDN to localize and isolate compromised devices. The SDN-based Smart Cyber Switching (SCS) framework for cyber restoration of a digital substation \cite{girdhar2025sdn}, which is itself an extension of an earlier work from the same authors \cite{girdhar2024sdn}, introduced SDN-based localization of malicious hosts. These SDN-based approaches benefit from complete network topology visibility, which enables straightforward port-level isolation actions once an alert has been raised. However, this approach builds upon the alerts generated by the IDS proposed in \cite{hong_integrated_2014} and therefore inherits its detection limitations. As a consequence, it would not be able to detect the cyber attacks considered in this paper, i.e., replay with spoofing and MitM attacks in High-availability Seamless Redundancy (HSR) rings. As a result, attack source localization based on this detection method would not work in such scenarios, leaving the digital substation exposed to undetected, unmitigated, and unlocalized MitM attacks. Other approaches presented in the literature that rely on SDN-based traffic monitoring, communication network throughput analysis, and attack graph models \cite{presekal2023attack} would not be able to reliably locate stealthy MitM attacks, since these attacks preserve the expected network throughput and traffic patterns observed by such methods. We quantitatively confirm both of these limitations in Section \ref{sec:localization_eval}, where the detection and localization algorithms underlying \cite{hong_integrated_2014} and \cite{presekal2023attack} are re-implemented and evaluated.

The method proposed in this paper differs from prior localization approaches in two fundamental architectural aspects. First, existing methods depend on SDN infrastructure, which is not available in our testbeds and is not currently deployed in the vast majority of operational digital substations. The proposed method, in contrast, operates over standard switched Ethernet networks and HSR rings, making it directly applicable to existing substation deployments without requiring infrastructure changes. Second, prior approaches localize the attacker by reading the SDN port to which the compromised host is connected from the controller’s topology map. Contrary, the proposed method infers the position of the compromised IED within an HSR ring from the spatial-temporal correlations of the Exponentially Modified Gaussian (EMG)-based timing fingerprint observed across multiple devices, without relying on any privileged topology information. While these architectural differences, together with the unavailability of SDN infrastructure, prevent a faithful reproduction of the complete prior frameworks, they do not preclude a comparison of the underlying detection and localization algorithms: as detailed in Section \ref{sec:localization_eval}, we re-implement the rule-based SV anomaly detection of \cite{hong_integrated_2014} and the throughput-based localization method of \cite{presekal2023attack} and evaluate them on the same dataset, data split, and metrics used for the proposed method.

In summary, the existing literature on intrusion detection, prevention, and attack source localization for IEC 61850 SV streams in digital substations exhibits four limitations: (i) measurement-based IDSs are vulnerable to replay attacks that are consistent with the system physical model; (ii) deep packet inspection-based IDSs are vulnerable to advanced MitM attacks that preserve $FS$, \texttt{smpCnt}, and SV stream identifier features; (iii) detection latencies reported in the literature, ranging from hundreds of microseconds to hundreds of milliseconds, are incompatible with the strict real-time constraints of IEC 61850-compliant intrusion prevention; and (iv) attack source localization methods proposed in the literature rely on SDN-based approaches that inherit the detection limitations of the underlying IDS and require an OT infrastructure that is not deployed in the majority of existing substations. Finally, none of the existing works exploits the information advantage provided by the statistical properties of the SV frame arrival time shift ($F_{as}$), combined with DL, to detect, prevent, and localize the source of SV-based injection attacks simultaneously, which is one of the contributions of this paper.

To address these gaps, this paper proposes an integrated method for detecting, preventing, and localizing the source of IEC 61850 SV cyber attacks in digital substations, aiming to enhance both situational awareness and operational security while meeting real-time communication requirements for protection schemes. 
The main contributions of this paper are summarized as follows:
\begin{enumerate}
  \item We present the first integrated method based on hybrid statistical-deep learning for real-time detection, prevention, and source localization of IEC 61850 SV cyber attacks in digital substations. At its core, the method introduces a novel analytical formulation of the SV frame arrival time statistics using EMG probability distributions, which capture network latency, jitter, and device clock drift during time synchronization loss. The combination of this statistical model with DL enables the detection and mitigation in real-time of MitM attacks and all known techniques of SV injection attacks.

  \item We employ Long Short-Term Memory (LSTM) models to capture temporal deviations in the estimated probability distributions, enabling timely detection of ongoing MitM attacks on SV streams against the protected IEDs.

  \item We adopt Elman RNNs to combine and correlate the estimated probability distributions across multiple IEDs within HSR rings. By analyzing the spatio-temporal correlations among Probability Distribution Functions (PDFs), the compromised device used to perform the MitM attack on SV streams is precisely localized from observable OT traffic alone, without relying on additional infrastructure such as SDN.
 
  \item We utilize the estimated PDFs to discern legitimate and malicious frames in real-time, allowing malicious frames to be discarded before affecting IED protection logic. This provides an effective intrusion prevention mechanism with minimal processing overhead and high operational explainability.
  
  \item We validate the proposed method for real-time detection, prevention, and source localization of IEC 61850 SV cyber attacks in digital substation on two lab testbeds and an industrial testbed at Elia transmission system operator in Belgium comprising industrial-grade protection and communication devices, hardware-in-the-loop simulations, virtualized IEDs (vIEDs) and merging units (vMUs), and high-fidelity emulated OT networks. Results demonstrate accurate detection and localization of MitM attacks on SV streams, reliable filtering of forged SV frames, a near-zero FPR in the absence of cyber attacks, negligible additional latency, and throughput exceeding 100,000 SV frames per second. The method remains resilient to communication latency, jitter, and time synchronization deviations, making it suitable for deployment in IEC 61850-based digital substations.
\end{enumerate}

The remainder of the paper is organized as follows. In Section II, background information about the SV protocol and cyber security concerns is provided. In Section III, the integrated method proposed for the detection, prevention, and source localization of IEC 61850 SV-based cyber attacks is presented. In Section IV, the proposed method is extensively validated, evaluated, and compared with the related state-of-the-art solutions. Finally, Section V concludes the paper.

\section{Background}
As introduced in the previous sections, the operation of digital substations relies on time-critical communication services defined within the IEC 61850 standard. Understanding both the operational principles of the SV protocol and its associated cyber security requirements is essential for assessing vulnerabilities and designing effective defence mechanisms. The following subsections first provide an overview of the IEC 61850 SV protocol and its deterministic timing characteristics, followed by a discussion on the security challenges and countermeasures proposed in the relevant IEC standards.

\subsection{IEC 61850 Sampled Values}
\label{sec:iec_61850_sv}
IEC 61850 SV is a publisher-subscriber protocol used for reporting three-phase current and voltage measurements from Merging Units (MUs) to IEDs. MUs publish SV frames in multicast, whereas IEDs process only the SV streams to which they are subscribed. The protocol was introduced in IEC 61850-9-2, further refined in IEC 61850-9-2LE, and ultimately standardized in IEC 61869-9. In digital substations, the most common SV frame publishing rates are 4000 and 4800 frames per second for 50 Hz and 60 Hz power systems, respectively. The standard also defines the structure of SV frames, including the \texttt{smpCnt} field, a counter that increments with each published frame and resets every second; the \texttt{smpCnt} value therefore ranges from zero to $FS-1$. This provides a certain level of determinism in the arrival time of each SV frame: for a given second $i$ and a \texttt{smpCnt} value $c$, the theoretical frame arrival time is
$F_a^e(i, c) = i + c/FS$, with  $c \in \{0, 1, \dots, FS - 1\}$.

This determinism across publishing periods provides an information advantage to discern legitimate and malicious SV frames, detect MitM attacks, and localize compromised IEDs within OT communication network. To further increase this determinism,  as recommended in IEC 61850-90-4, priority tags and proper communication network engineering should be used to ensure the lowest possible latency and jitter for SV frames \cite{zhou_propagation_2022}.

\subsection{Sampled Values Cyber Security} \label{sec:sv_cybersec}

Considering the strict latency requirements of power system protection schemes of 3 ms, the IEC 62351 standard stipulates against the application of digital signatures due to increased latency and processing times. To counteract most of the possible cyber attacks exploiting the SV protocol, the standard recommends the usage of Message Authentication Codes (MAC) relying on Hash-based MAC (HMAC) or Advanced Encryption Standard-Galois MAC (AES-GMAC) to provide message authentication. While appending the computed MAC to published frames helps mitigate replay, spoofing, and MitM attacks, it also introduces multiple challenges and limitations. These include the increased publishing and subscribing devices computational load, communication latency, and message overhead, the need for Public Key Infrastructure (PKI) and multiple Key Distribution Centers (KDCs) deployment, lack of support for legacy devices not supporting IEC 62351-6, and the risk of pre-shared keys violation. Moreover, each device belonging to the same Group Domain of Interpretation (GDOI) has access to the same shared secret key, and thus can perform spoofing attacks. 

Summarizing, once an attacker is able to gain access to the digital substation OT communication network \cite{christopher2024sans}, the cyber attacks that can be performed against an SV subscriber include: (1) flooding, where a large number of SV frames are injected into the communication network by the attacker to exhaust subscriber resources; (2) spoofing, where the attacker injects malicious SV frames pretending of being the legitimate SV publisher; (3) replay, where the attacker re-injects into the communication network the previously sniffed SV frames; (4) high \texttt{smpCnt} attack, in which an attacker injects one or multiple SV frames with a high \texttt{smpCnt} field value, causing the SV subscriber implementing the replay protection mechanism as mandated in IEC 62351-6 to discard any further legitimate SV frame with a lower \texttt{smpCnt}; (5) MitM, where an attacker takes control of a device forwarding SV frames from the publisher to the subscriber and tampers with the forwarded SV frames; (6) access to the pre-shared secret key, in which the attacker gains access to the pre-shared secret key used for MAC generation and validation and injects authenticated malicious SV frames. 

In this paper, to evaluate the effectiveness of the proposed method, the performed cyber attacks assume the injection of malicious SV frames that are completely indistinguishable from the legitimate ones. In other words, the Hamming distance between the content of legitimate and malicious SV frames is equal to zero. By proving the method's effectiveness in this worst-case scenario, the results can be extended to cases where false measurements are injected, thus leading to a Hamming distance greater than 0.

\section{Cyber Attacks Detection, Prevention, and Source Localization}

\begin{figure}[!t]
\centering
\includegraphics[width=\linewidth]{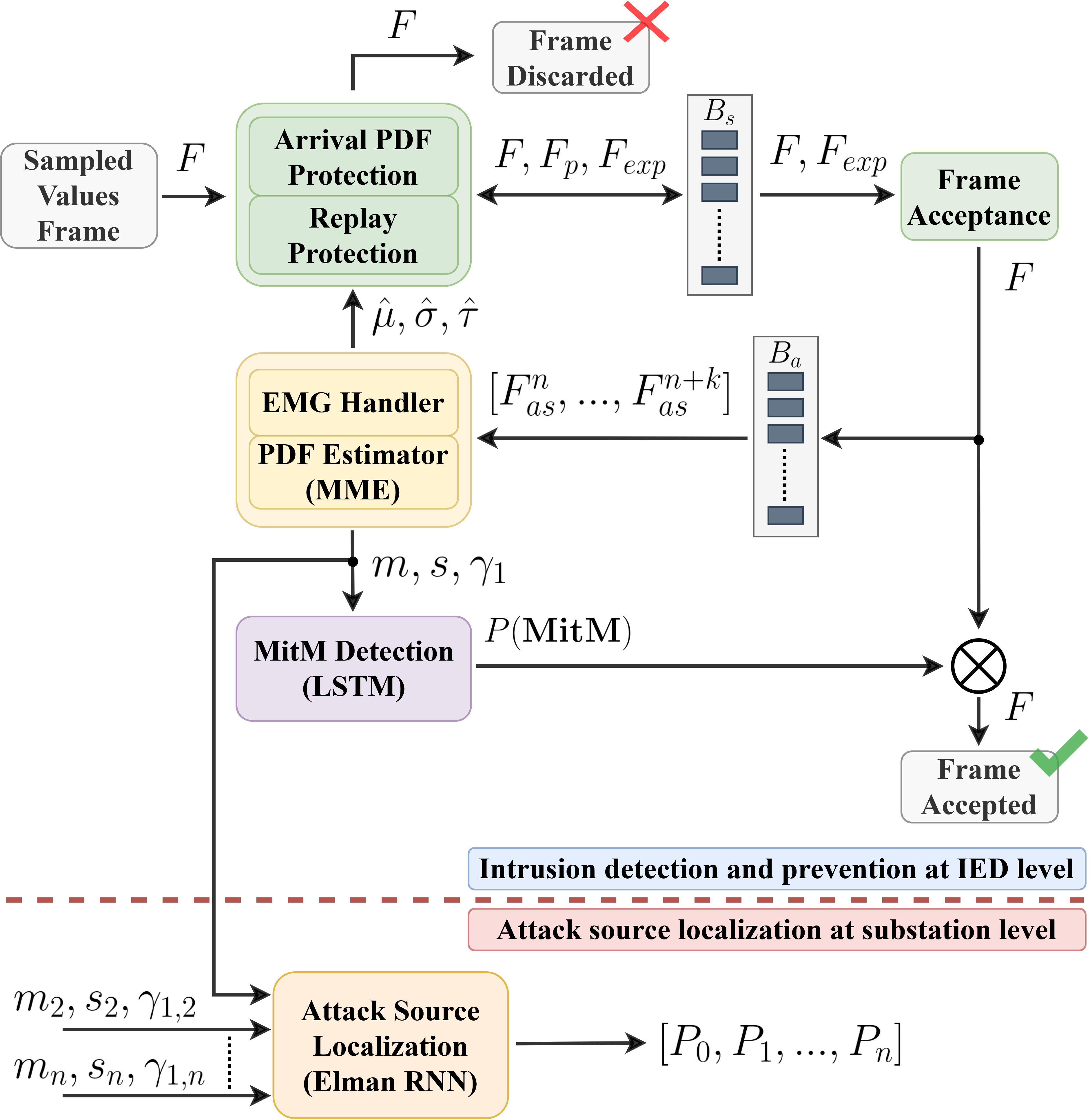}
\caption{Hybrid statistical-deep learning-based intrusion detection, prevention, and attack source localization system architecture.}
\label{fig_1}
\end{figure}

The proposed integrated method can be divided into two main modules, i.e., (1) the intrusion detection and prevention module, which is deployed in each IED, and (2) the attack source localization module, which is deployed on a centralized location within the digital substation and receives statistical information from each IED. Fig. \ref{fig_1} represents the architecture of the proposed method. In the following, theoretical preliminaries and the different components and their interactions are presented.

\subsection{SV Frames Arrival Time Shift}
\label{sec:sv_frames_fas}
In real communication networks, the theoretical frame arrival time defined in Section \ref{sec:iec_61850_sv} is affected by frame transmission, propagation, queuing, and processing delays, as well as by time synchronization errors among SV stream publisher and subscriber devices. As a result, each frame is received at a time $F_a^{r} (c)\neq F_a^e (c)$ within each second. By monitoring the SV subscriber's Network Interface Card (NIC), $F_a^{r} (c)$ can be measured, and the frame arrival time shift $F_{as}$, defined as the difference $F_a^{r}(c) - F_a^e(c)$, can be computed.

To account for the stochasticity of the communication network infrastructure, a PDF can be estimated over $F_{as}$. Then, the modeled PDF can be used as an information advantage for intrusion detection, prevention, and attack source localization. Indeed, once this PDF is estimated, a legitimacy probability can be assigned to each received frame, allowing the distinction between legitimate and malicious SV frames based on their $F_{as}$ measured at the subscriber. Regarding the choice of the most appropriate PDF to be used, the literature shows that latency in communication networks is often modeled using an exponential distribution \cite{benvenuto2011principles}. However, the exponential distribution assigns a probability equal to zero to samples below its lower bound, making it unsuitable for scenarios where latency reductions may occur due to device time-synchronization errors. To address this limitation, the EMG distribution is used to model $F_{as}$. The EMG, defined as the sum of a Gaussian and an exponential component, effectively represents both the narrow, near-deterministic base communication delay and the heavy-tailed positive jitter caused by queueing, contention, and background traffic. Such characteristics enable the EMG to capture the sharp mode and long right tail observed in empirical network latency distributions more accurately than either distribution alone.

Finally, it should be remarked that, while related, the inter-frame arrival time, i.e., I-FAT, and SV frames arrival time shift, i.e., $F_{as}$, differ fundamentally in definition, probability distribution, information content, and susceptibility to communication network latency and jitter. In fact, I-FAT represents the time difference between consecutive SV frames and follows a Gaussian distribution centered at $1/FS$, whereas $F_{as}$ measures the deviation between the expected and actual arrival times for individual frames, modeled by an EMG distribution whose mean reflects communication network latency and clock synchronization drift \cite{kim2022industrial}. These distinctions represent a fundamental difference compared to how the SV frame arrival times information has been exploited in existing studies.

\subsection{Exponentially Modified Gaussian Distribution}
\label{sec:emg_distribution}

The EMG distribution is a probability distribution whose PDF is obtained by convoluting a normal distribution, $Z \sim \mathcal{N}(\mu, \sigma^2)$, with an exponential one, $E \sim \mathrm{Exp}(\lambda)$. An EMG distribution is characterized by the following PDF:

\begin{equation}
\label{deqn_ex3}
f_{EMG}(x) = \frac{\lambda}{2} e^{\frac{\lambda}{2} (2\mu + \lambda \sigma^2 - 2x)} \, \mathrm{erfc}\left( \frac{\mu + \lambda \sigma^2 - x}{\sqrt{2} \, \sigma} \right)
\end{equation}

\noindent where $\mu$ and $\sigma$ are the mean and standard deviation of the normal distribution, respectively, and $\lambda = {1}/\tau$ is the exponential decay of the exponential distribution. Due to the changing operating conditions, $F_{as}$ can vary in time, and thus EMG distribution parameters need to be continuously estimated and updated using the latest received SV frames. As shown by Ali et al. \cite{ali_comparison_2022}, the Method of Moments Estimation (MME) provides the best trade-off in terms of parameters estimation accuracy and efficiency. In fact, to minimize SV frames classification latency, an accurate estimator providing a closed form solution is required. By using the MME, it is possible to equate the first three theoretical moments of the EMG distribution with the three empirical ones as in (\ref{deqn_ex4}), (\ref{deqn_ex5}), and (\ref{deqn_ex6}), for the mean ($m$), variance ($s^2$), and skewness ($\gamma_1$), respectively:
\begin{equation}
\label{deqn_ex4}
\mathbb{E}[X] = m = \mu + \tau = \hat{m}_1
\end{equation}
\begin{equation}
\label{deqn_ex5}
\mathrm{Var}(X) = s^2 = \sigma^2 + \tau^2 = \hat{m}_2
\end{equation}
\begin{equation}
\label{deqn_ex6}
\gamma_1 = \frac{2\tau^3}{\left(\sigma^2 + \tau^2\right)^{3/2}} = \frac{\hat{m}_3}{\hat{m}_2^{3/2}}
\end{equation}

\noindent where $\hat{m}_r$ is the $r-th$ empirical moment computed on the observed data. As shown in Oliver et al. \cite{olivier_positively_2010}, solving for the EMG distribution parameters gives:

\begin{equation}
\label{deqn_ex7}
\hat{\mu} = m - s \left( \frac{\gamma_1}{2} \right)^{1/3}
\end{equation}
\begin{equation}
\label{deqn_ex8}
\hat{\sigma}^2 = s^2 \left[ 1 - \left( \frac{\gamma_1}{2} \right)^{2/3} \right]
\end{equation}
\begin{equation}
\label{deqn_ex9}
\hat{\tau} = \frac{1}{\hat{\lambda}} = s \left( \frac{\gamma_1}{2} \right)^{1/3}
\end{equation}

\noindent where $\hat{\mu}$, $\hat{\sigma}$, and $\hat{\lambda}$ are the estimated PDF parameters.

\subsection{Intrusion Detection and Prevention}
To discern legitimate and malicious SV frames on a per-frame basis in real-time, and to detect MitM attacks, the proposed method relies on four interdependent components being executed in parallel in each IED. Moreover, two buffers, i.e., the staging buffer ($B_s$) and the accepted frames buffer ($B_a$), are used to store the latest received frames which are most likely to be legitimate, and the last $k$ accepted frames, respectively.

\subsubsection{Arrival PDF and Replay Protections}

Once a new SV frame is received at the subscriber NIC, it is processed by the first component. First, it is checked whether frames with the same \texttt{smpCnt} value were already accepted in the current time period. If it is the case, the replay protection mechanism is triggered, and the frame is discarded. Then, after computing $F_{as}$ for the current frame as presented in the previous section, the EMG PDF ($f_{EMG}$) is used to compute the probability of the frame being legitimate ($F_p$), such that $F_p = f_{EMG}(F_{as})$. Subsequently, the component checks whether in $B_s$ a frame with the same \texttt{smpCnt} value as the just received frame is present. If such a frame is already present, but the probability of it being legitimate is higher than the one computed on the last received frame, the last received frame is discarded. Otherwise, if no frame with the same \texttt{smpCnt} is present in $B_s$, or the legitimacy likelihood of the new frame is higher, the just received frame is inserted in the buffer, and the already present one is discarded. In this way, it is ensured that exactly $FS$ frames per second are accepted and no false positives can occur when no cyber attack is ongoing. Other than computing $F_p$, a frame expiration time ($F_{exp}$) is assigned to each frame added to $B_s$, with $F_{exp}$ computed as follows:
\begin{equation}
\label{deqn_ex10}
F_{exp} = 
\begin{cases}
0, & F_{as} \geq \mathbb{E}[F_{as}]^t \\
F_a + r, & F_p \geq f_{EMG}(\mathbb{E}[F_{as}]^t) \\
F_a + r + 3 \cdot \hat{\sigma}^t, & \text{otherwise}
\end{cases}
\end{equation}
where the residual $r$ is equal to $\mathbb{E}[F_{as}]^t - F_{as}$.

\begin{figure}[!b]
\centering
\includegraphics[width=\linewidth]{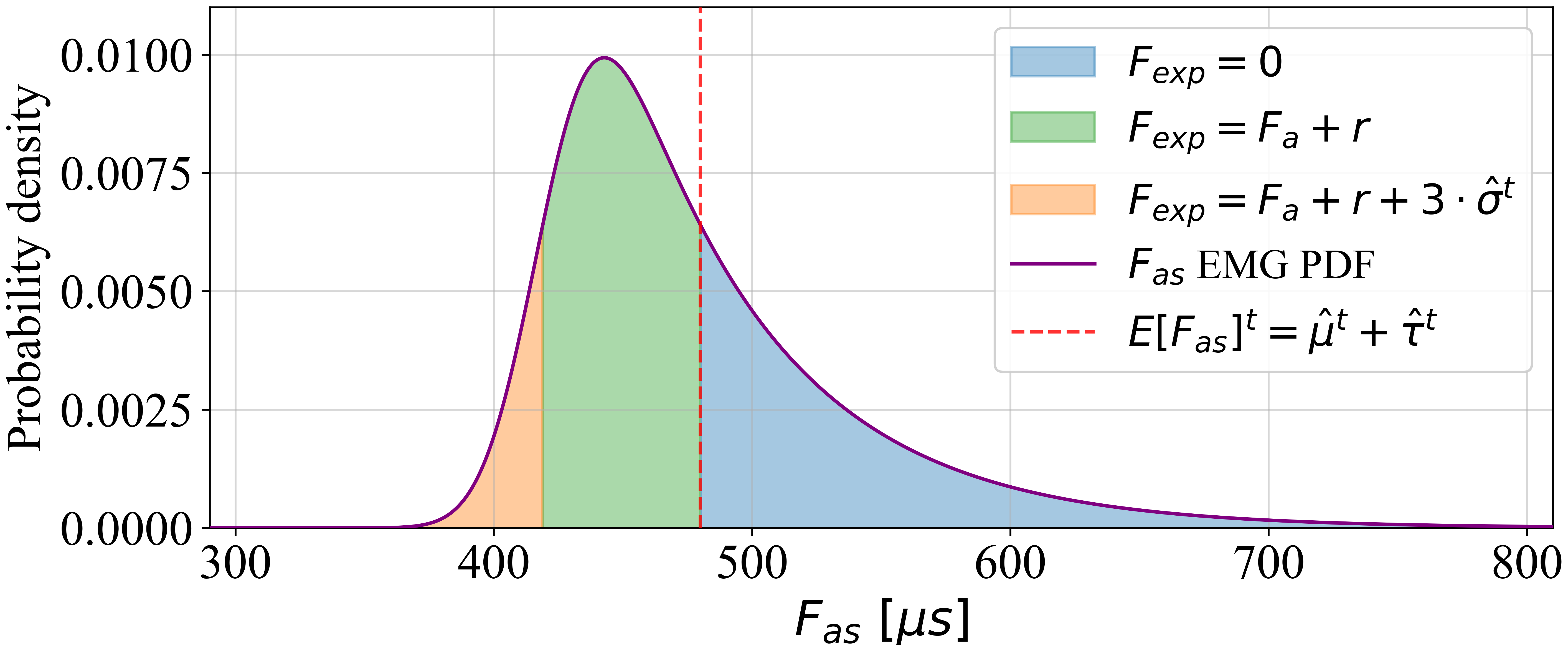}
\caption{Graphical interpretation of $F_{exp}$ value assignment.}
\label{fig_2}
\end{figure}

The rationale behind the values assigned to $F_{exp}$ is to minimize the time frames remain buffered in $B_s$ while ensuring that frames with the highest likelihood of originating from the legitimate device, rather than from the attacker, are accepted. In fact, if a frame with \texttt{smpCnt} equal to $c$ is received at time $F_a'$, and the computed $F_{as}'$ is equal to or greater than $\hat{\mu}^t + \hat{\tau}^t = \mathbb{E}[F_{as}]^t$, then, for any other frame with the same \texttt{smpCnt} received at a time $F_{as}'' > F_{as}'$, $f_{EMG}(F_{as}'') = F_p''$ will be lower than $f_{EMG}(F_{as}')  = F_p'$. Thus, the frame can be accepted immediately without any further delay. In the other cases, it cannot be inferred that frames with higher legitimacy likelihood will not be received in the future, and thus the received frame needs to be retained until $F_{exp}$. A graphical interpretation on the values assigned to $F_{exp}$ is provided in Fig. \ref{fig_2}. Moreover, the pseudocode for this component is provided in Algorithm \ref{alg:alg1}.

\subsubsection{Frame Acceptance}

This component is responsible for processing SV frames stored in $B_s$, while continuously monitoring whether the current time is higher than frames’ $F_{exp}$. If this condition holds, it implies that no additional SV frames with the same \texttt{smpCnt} field value and a higher likelihood of legitimacy are expected to be received. Thus, the SV frame currently stored in $B_s$ can be accepted, forwarded to the protection scheme for immediate use, and added to $B_a$, which stores the SV frames used for updating the parameters of the EMG PDF.

\begin{algorithm}[!t]
    \caption{SV frame likelihood ($F_p$) and expiration ($F_{exp}$) computation, and push frame to staging buffer.}
    \label{alg:alg1}
    \DontPrintSemicolon
    \SetKwInOut{Input}{Input}
    \SetKwInOut{Output}{Output}
    \SetKw{Discard}{Discard}
    \SetKw{Return}{Return}

    \newcommand{\Exp}[1]{\mathbb{E}[#1]^t}
    \newcommand{\Sig}{\hat{\sigma}^t}
    \newcommand{\Femg}{f_{\text{EMG}}}
    
    \Input{SV frame $F$, Frame arrival time shift ($F_{as}$), Current buffer $B_s$}
    \Output{Updated buffer $B_s$ or Discard frame}
    
    \If{$| F_{as} - \Exp{F_{as}} | \geq 3 \cdot \Sig$}{
        \Discard $F$ \; 
        \Return
    }

    $F_{p} \gets \Femg(F_{as})$ \;
    
    \uIf{$F_{as} \geq \Exp{F_{as}}$}{
        $F_{exp} \gets 0$ 
    }
    \uElseIf{$F_{p} \geq \Femg(\Exp{F_{as}})$}{
        $F_{exp} \gets F_a + (\Exp{F_{as}} - F_{as})$ \;
    }
    \Else{
        $F_{exp} \gets F_a + (\Exp{F_{as}} - F_{as}) + 3 \cdot \Sig$ \;
    }

    $F_{new} \gets \{F, F_p, F_{exp}\}$ \;
    \eIf{$\exists F' \in B_s \text{ s.t. } F'_{smpCnt} == F_{smpCnt}$}{
        \If{$F'.F_p < F_p$}{
            Replace $F'$ with $F_{new}$ in $B_s$ \;
        }
        \Else{
            \Discard $F$ 
        }
    }{
        Insert $F_{new}$ into $B_s$ maintaining order by $smpCnt$ \;
    }
    
    \Return
\end{algorithm}

\subsubsection{EMG Handler and PDF Estimator}
\label{sec:emg_handler_and_pdf_estimator}

After the number of SV frames in $B_a$ is greater than $k$, this component takes care of estimating and updating the parameters of the EMG distribution. This is done by first updating the samples’ moments with a weighted mean between the previously calculated moments ($\hat{m}^t_r$) and the just computed ones ($\underline{\hat{m}_r}$), as in (\ref{deqn_ex11}), and then using the updated moments values to compute the updated EMG parameters  $\hat{\mu}^{t+1}$, $\hat{\sigma}^{t+1}$, and $\hat{\lambda}^{t+1}$.
\begin{equation}
\label{deqn_ex11}
\hat{m}_r^{t+1} = \frac{FS \cdot \hat{m}_r^{t} + k \cdot \underline{\hat{m}}_r}{FS + k}
\end{equation}

\noindent The updated EMG parameters are then used by the arrival PDF and replay protections component to evaluate the legitimacy of any SV frame that arrived subsequently.

The choice of parameter $k$ directly affects the trade-off between adaptation speed and statistical stability of the EMG parameters estimation. Smaller values of $k$ allow the method to adapt faster to changing network operating conditions but produce noisier parameter estimates due to the limited sample size. Conversely, larger values of $k$ yield more stable parameter estimates but slow the adaptation to changes in network behavior. Furthermore, the time required to compute the moments scales linearly with the number of samples, and the proposed method is designed to distribute this computational cost across the sampling period rather than concentrating it in a single bulk operation over a large sample. The value of $k$ was therefore set to 400, the smallest sample size yielding stable parameter estimates across all evaluated datasets. This choice ensures numerical stability while preserving the ability to track variations in network operating conditions in near real-time and to maintain a uniform computational load over time. The empirical justification of this choice is presented in Section \ref{sec:fas_characterization}.

\subsubsection{MitM Detection}
Whereas the first component is meant to accept exactly $FS$ SV frames per second while minimizing the likelihood of malicious frames being accepted, this fourth component aims at detecting MitM attacks on SV streams. In this type of attack, the attacker can completely block the legitimate SV stream and inject a malicious one by compromising one forwarding device between the SV publisher and subscriber. Thus, the subscriber device receives exactly $FS$ frames per second as expected. Still, as experimentally shown, the act of intercepting and tampering with the forwarded SV frames causes an inevitable change in the latency and jitter statistics of the received SV stream at the subscriber level. In fact, to perform the MitM attack, the attacker is required to change the NIC configuration of the compromised forwarding device to modify the frames forwarding rules or to disable the HSR bridge between the two NICs. Then, the network traffic needs to be monitored, tampered with, and forwarded from one NIC to the other in user-space level on the device, causing the inevitable alteration of the forwarded traffic statistical properties. A RNN is deployed to detect anomalous variations in the EMG parameters. The input of the RNN consists of the mean ($m$), standard deviation ($s$), and skewness ($\gamma_1$) of the $F_{as}$ measured on the accepted frames, calculated over a sliding window of 200 samples with a step size of 50 samples. The RNN outputs the likelihood of an ongoing MitM attack. A grid search approach is used to select the most appropriate RNN model and perform hyperparameter optimization. The considered RNN models include Elman RNN, LSTM, and Gated Recurrent Units (GRU), whereas the dimension and the number of hidden layers vary from 3 to 20 and from 2 to 8, respectively. The model providing the best trade-off between model complexity, detection accuracy, and prediction time consists of 4 stacked LSTM cells of 20 neurons each. Finally, to perform the binary classification, a fully connected layer is stacked to the output of the last LSTM cell.

\subsection{Attack Source Localization}

As mentioned in the previous section, anomalous changes in the PDF of $F_{as}$ can be an indicator of an ongoing MitM attack. By aggregating and correlating the statistical properties of the SV frames arrival times from different IEDs deployed within the digital substation, it is possible to identify and locate which device is performing a MitM attack. In this paper, the focus is on IEDs deployed in an HSR ring topology. As represented in Fig. \ref{fig_1}, the mean ($m_i$), standard deviation ($s_i$), and skewness ($\gamma_{1,i}$) of the measured $F_{as}$ from each $i-th$ IED are delivered to the server hosting the attack source localization module within the digital substation. These metrics are then used to estimate the probability of a MitM attack being ongoing, and the probability for each IED of acting maliciously. As for the MitM component, a sliding window of 200 $F_{as}$ samples with a step size of 50 samples is used. This last component consists of 5 stacked Elman RNN cells; each cell hidden layer contains 26 neurons. The tanh activation function is used. The output of the last RNN cell is provided as input to a fully connected layer to perform the final multi-class classification. To improve the DL model accuracy and increase its generalization capabilities, the component’s input data is derived from the statistical metrics received from each IED as follows:
\[
\forall t \in \{1,\dots,T\},\forall i,j \in \{1,\dots,N\},\, i<j:
\]
\begin{equation}
\label{deqn_ex12}
\begin{cases}
m_{i,j}^t = m_i^t - m_j^t \\
s_{i,j}^t = s_i^t - s_j^t \\
\gamma_{1,i,j}^t = \gamma_{1,i}^t - \gamma_{1,j}^t
\end{cases}
\end{equation}

\noindent where $t$ is the current time window, $N$ is the number of IEDs deployed in the HSR ring, and $m_n^t$, $s_n^t$, and $\gamma_{1,n}^t$ are the mean, standard deviation, and skewness measured at $n-th$ IED at time $t$. This data transformation prevents the DL model from overfitting the absolute values measured in the specific communication network conditions, but to focus on the relative evolution of the statistical properties.

\subsection{Attacker Considerations} \label{sec:attacker_considerations}
From an attacker's perspective, several challenges need to be overcome to deceive the proposed intrusion detection, prevention, and localization method. First, estimating $F_{as}$ for a specific SV subscriber requires accurate knowledge of the overall communication latency and jitter between the legitimate publisher and the targeted subscriber, and between the compromised device and the targeted subscriber, while also accounting for timing synchronization error. In fact, accurate estimation of the EMG parameters can only be performed on the devices already compromised by the attacker by having access to real-time traffic observations; the EMG parameters of non-compromised devices are not observable, and their accurate estimation is impractical. Second, to inject SV frames at the exact time instants required to match the expected $F_{as}$ statistical properties, the compromised device must run a real-time Operating System (OS) and be time synchronized via Precision Time Protocol (PTP). However, even under these conditions, the CPU scheduler and the OS kernel network stack introduce stochastic processing delays when processing and re-transmitting the forwarded, tampered SV frames. Even assuming that the compromised device is running a real-time OS, predicting and overcoming these stochastic delays to inject the malicious SV frames and match the estimated EMG PDF would be challenging. Dedicated hardware, such as FPGA-based packet processing, would be required to overcome these challenges, thus significantly increasing the cost and complexity of the attack. Third, a malicious script capable of publishing SV frames must be delivered, installed, and executed with high privileges on the compromised device. This deployment must occur stealthily, avoiding detection by both host-based and network-based IDSs, and without exhausting the compromised device's resources. Fourth, EMG parameters drift continuously with network load and clock synchronization conditions. The proposed DL-based attack detection and localization method exploits these temporal dynamics by identifying anomalous variations in the sequence of estimated PDFs over time, requiring an adversary to track and reproduce not just the instantaneous distribution but its full temporal evolution in real-time, constituting an additional and continuously shifting estimation problem. Finally, to cause protection schemes to react to a fault, it is not sufficient to have only a small number of false measurements to be accepted. For instance, when overcurrent protection is implemented, the measured current needs to be above a certain threshold for at least 200 ms. This means that an attacker needs at least $FS/5$ consecutive malicious frames to be accepted to trigger the protection scheme. Thus, the injected malicious frames need to be consistently accepted for a non-negligible amount of time. 
The combination of these conditions imposes considerable complexity on potential attacks, effectively preventing successful deception of the proposed method, even under ideal scenarios. In the remainder of the paper, it is assumed that the attacker has gained privileged access to a machine connected to the network switch forwarding the legitimate SV stream, or to a forwarding device itself when considering MitM attacks, and can execute arbitrary scripts with high privileges to directly access the compromised machine's NIC and inject malicious SV frames.

\begin{figure*}[!b]
  \centering
  \subfloat[Three vIED in a HIL configuration and subscribing to the SV stream published by RTDS through the GTNETx2 card.]{%
    \begin{minipage}[c][0.25\linewidth][c]{0.7\columnwidth}
      \centering
      \includegraphics[width=\columnwidth]{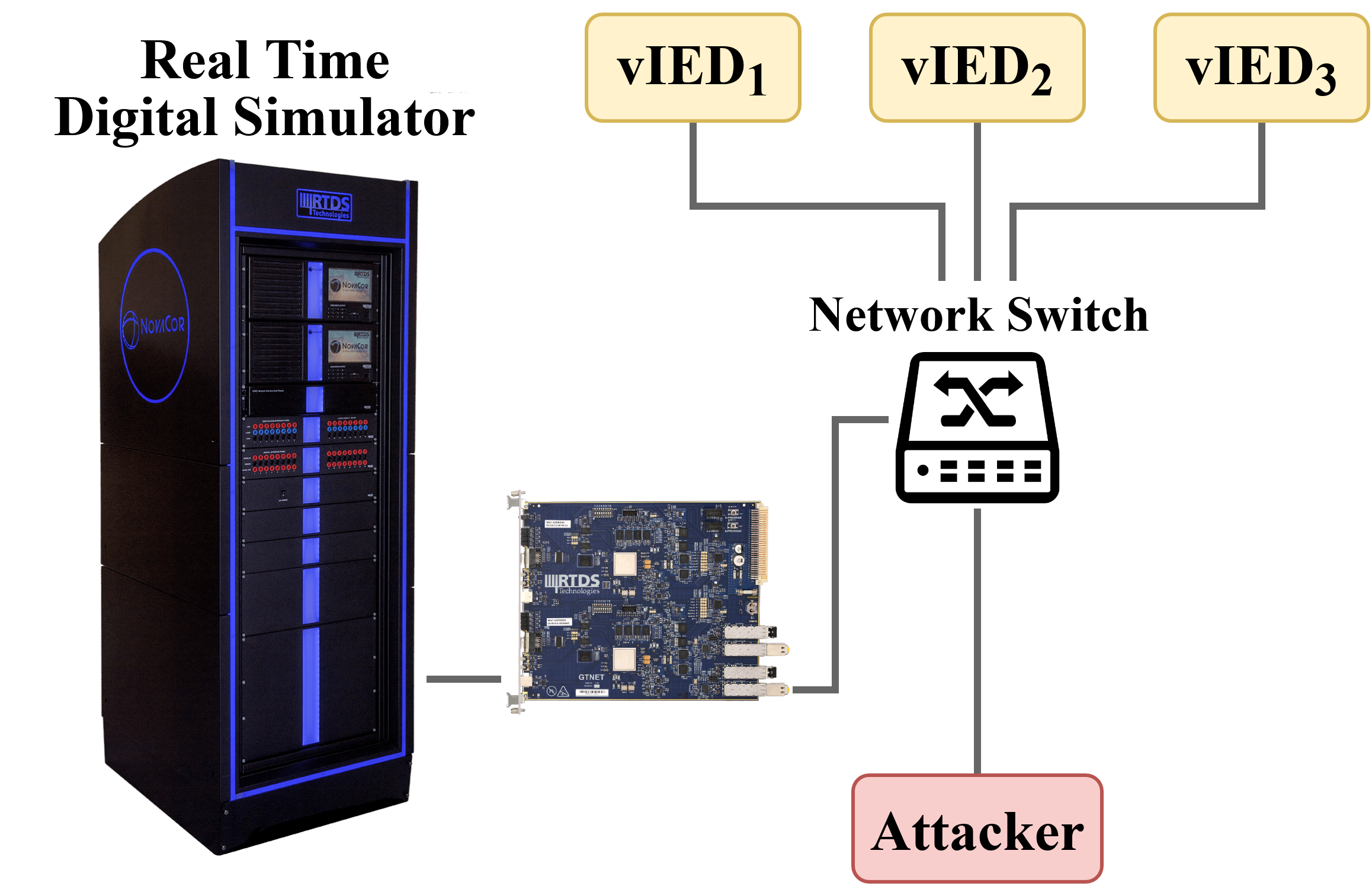}%
    \end{minipage}%
  }\hfill
  \subfloat[Industrial MU, IED, and network switches in PRP configuration.]{%
    \begin{minipage}[c][0.25\linewidth][c]{0.62\columnwidth}
      \centering
      \includegraphics[width=0.85\columnwidth]{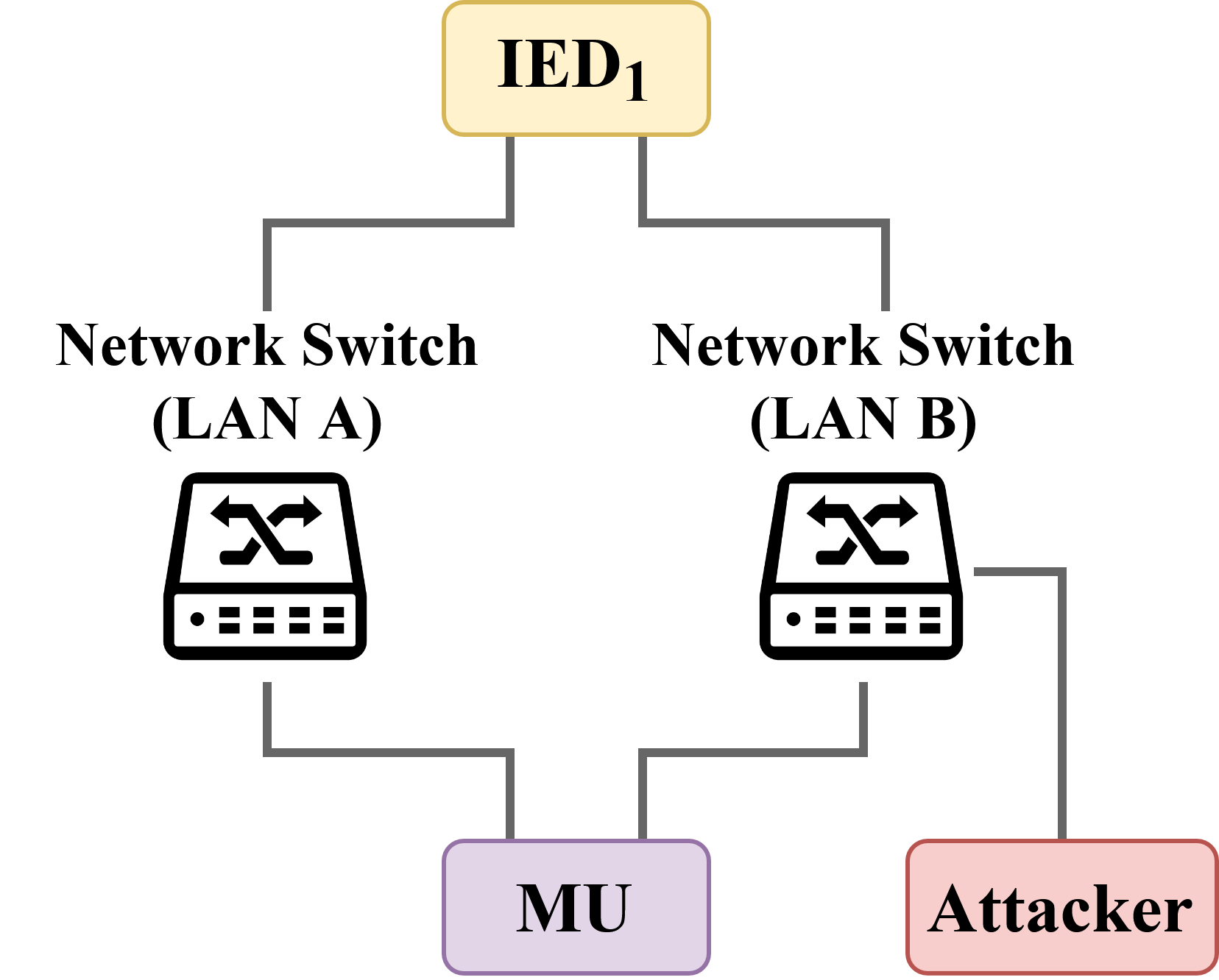}%
    \end{minipage}%
  }\hfill
  \subfloat[Fully virtualized setup with vMUs, vIEDs, and QuadBox deployed in Mininet; the third vIED is shown as a compromised device for illustration purposes.]{%
    \begin{minipage}[c][0.25\linewidth][c]{0.62\columnwidth}
      \centering
      \includegraphics[width=0.81\columnwidth]{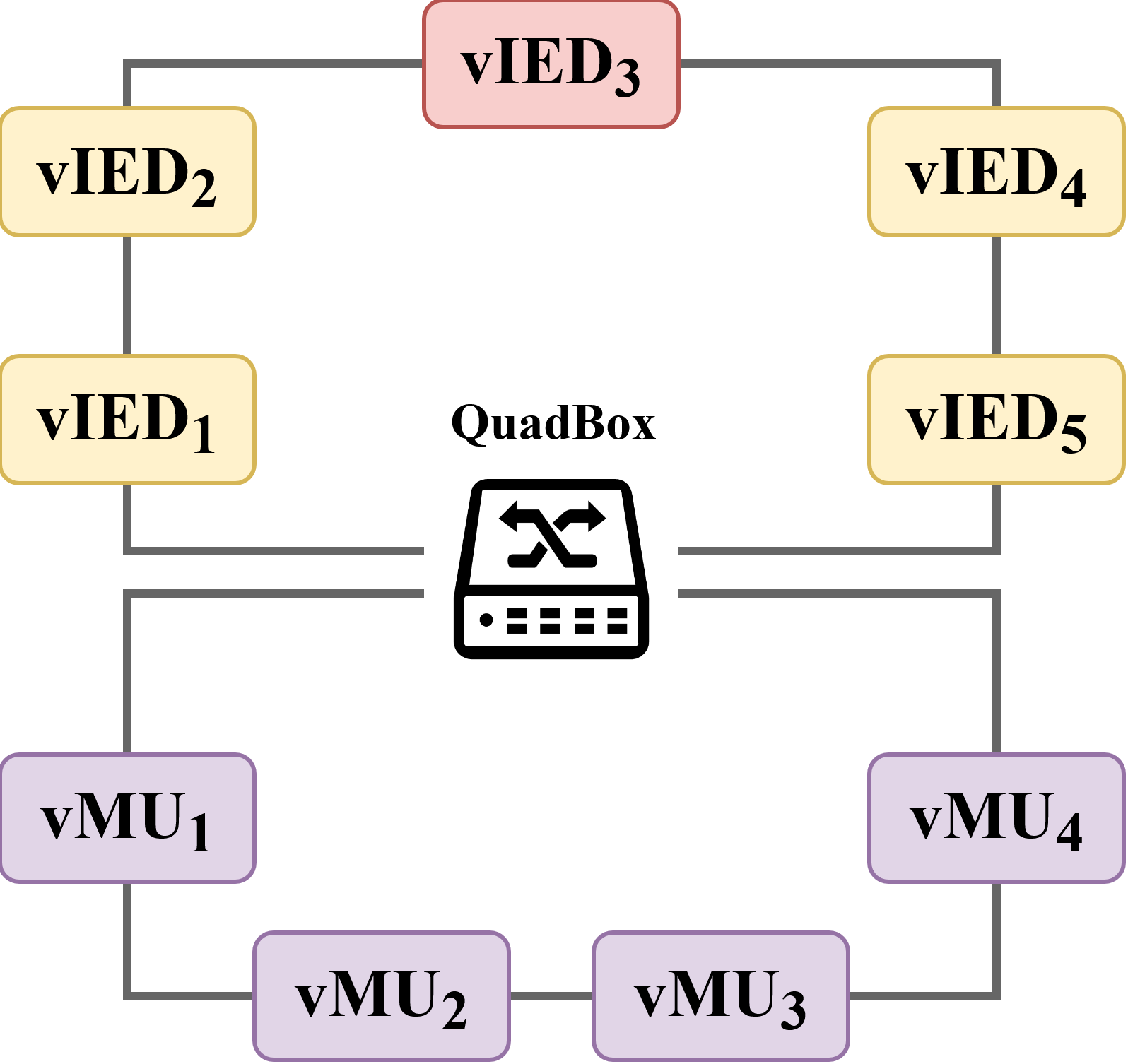}%
    \end{minipage}%
  }
  \caption{Schematic overview of the three experimental testbeds used to validate the proposed method.}
  \label{fig:testbeds}
\end{figure*}

\section{Experimental Results and Discussion}

The proposed method was extensively tested on data acquired from three different IEC 61850-compliant testbeds. The first testbed is deployed at Delft University of Technology. A Real-Time Digital Simulator (RTDS) runs an electromagnetic transient simulation of the IEEE 5-bus system in real-time. The three-phase current and voltage measurements generated in RTDS are sent to a network switch via a GTNETx2 card that acts as an SV publisher at 4800 frames per second. The GTNETx2 card also acts as a GOOSE subscriber to receive the trip signals sent by the three vIEDs. The three vIEDs are connected to the network switch in a star topology, and each of them is subscribed to the SV stream published by RTDS. The vIEDs run on general-purpose Virtual Machines (VMs) equipped with a 4-core CPU @ 3.6 GHz, 16 GB of RAM, running Ubuntu 22.04 OS, and relying on an open-source IEC 61850 C library \cite{zillgith_libiec61850}. The second testbed is deployed at Elia, the Belgian transmission operator, and represents an industrial-grade installation operating in a Parallel Redundancy Protocol (PRP) configuration. The testbed comprises an industrial MU publishing an SV stream at 4000 frames per second over two redundant local area network segments, two IEC 61850-compliant network switches, and an industrial IED acting as the SV subscriber. The third testbed consists of a fully virtualized setup in which the digital substation communication network is emulated in Mininet on a host machine equipped with an 8-core CPU @ 3.6 GHz, 32 GB of RAM, and running Ubuntu 22.04 OS. The setup comprises two HSR rings connected through a QuadBox, hosting 4 vMUs and up to 12 vIEDs, respectively. Each vMU publishes an SV stream at 4000 frames per second, and each vIED subscribes to one or more SV streams depending on the experimental scenario. The vMU and vIEDs are implemented using the same open-source IEC 61850 C library as in the first testbed. The architecture of the three testbeds is depicted in Fig. \ref{fig:testbeds}.

The experiments in the first testbed involved both SV injection attacks from a compromised device connected to the network switch and MitM attacks against one vIED. The second testbed focused exclusively on SV injection attacks from a compromised device connected to one of the network switches. Finally, the third testbed was dedicated to performing MitM attacks from each vIED in the HSR ring, one at a time, with a varying number of deployed vIEDs. During SV injection attacks, the attacker injected malicious measurements while spoofing the identity of the legitimate SV publisher, emulating persistent false fault conditions that triggered the (v)IEDs to incorrectly issue trip commands for opening circuit breakers. The IPS aimed to block these malicious SV frames from reaching the protection scheme while minimizing the drop of legitimate traffic. MitM attacks against the vIEDs verified that changing NIC configuration and forwarding and tampering SV frames in user-space through a highly optimized C script would still cause detectable changes in the statistical distribution of frame arrival times received by the targeted IED. Finally, data collected in the third testbed was used for training and testing the attack source localization module. In all cases, network traffic acquisitions were obtained by running Wireshark on each (v)IED and network switch across the three testbeds. Notably, the integrated method classified frames, detected MitM attack, and localized the compromised devices without inspecting measurement data, rendering differences between legitimate and malicious measurements irrelevant for this evaluation.

\subsection{Frame Arrival Time Characterization}
\label{sec:fas_characterization}
\begin{figure}[!b]
\centering
\includegraphics[width=\linewidth]{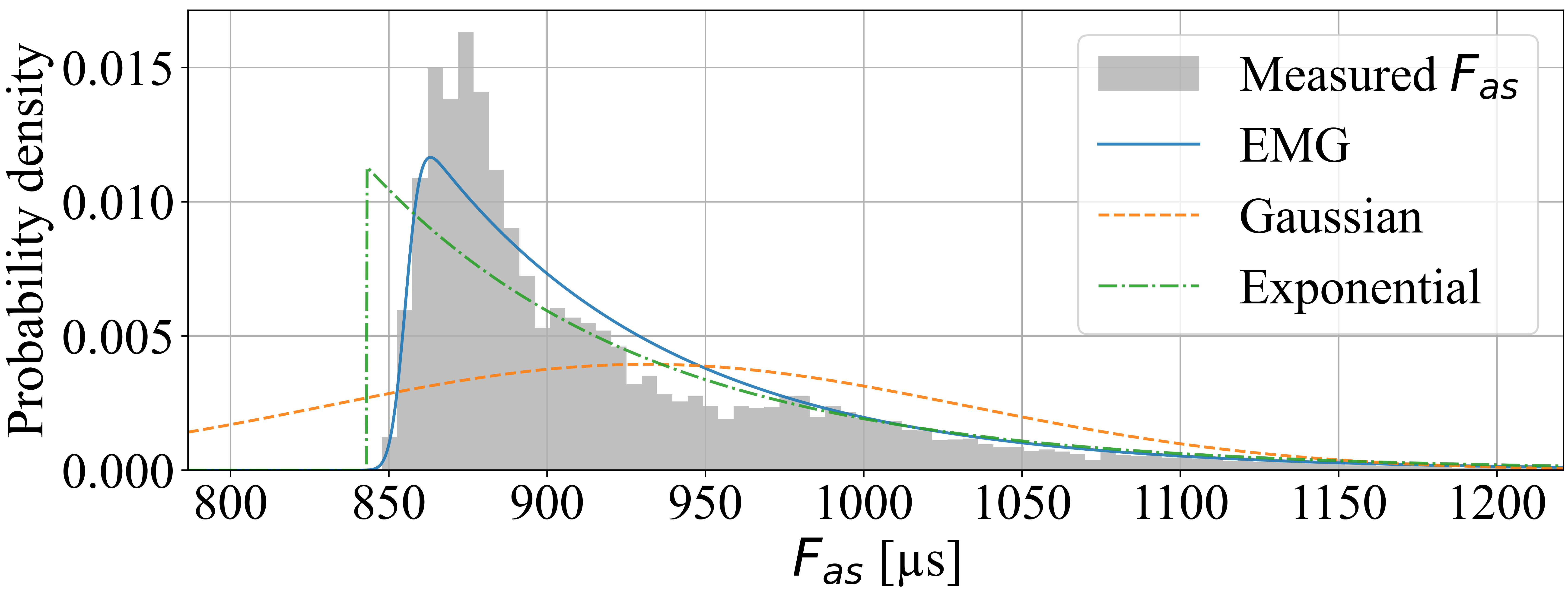}
\caption{Comparison of EMG, Gaussian, and exponential distribution fits to the measured $F_{as}$.}
\label{fig_3}
\end{figure}

The main information about the monitored SV streams in the first two testbeds during normal conditions is reported in Table \ref{tab:sv_streams_info}. This information includes the number of frames received per second ($FS$), measured $F_{as}$ mean and standard deviation, and the overall time synchronization error due to the lack of time synchronization with the PTP master clock. It should be noted that to comply with IEC 61850-9-3, each IED should guarantee a maximum holdover of ± 0.2 $\mu s/s$, thus implying a maximum time synchronization divergence among two devices of ± 0.4 $\mu s/s$. Given the holdovers observed during our tests, it is reasonable to assert that the proposed method was evaluated under the most challenging conditions likely to be encountered in real digital substation environments. Fig. \ref{fig_3} illustrates how the EMG distribution aligns with the measured $F_{as}$ values at one vIED deployed in the third testbed. 

\begin{table}[!b]
\caption{Information about the monitored SV streams at different (v)IEDs.}
\label{tab:sv_streams_info}
\centering
    \begin{tabular}{lcccc}
        \hline
        \textbf{} & vIED\textsubscript{1} & vIED\textsubscript{2} & vIED\textsubscript{3} & IED\textsubscript{1} \\
        \hline
        $FS$ [$frames/s$] & 4800 & 4800 & 4800 & 4000 \\
        $\mathbb{E}[F_{as}]$ [$ms$] & -7.08 & -4.60 & 0.447 & -113.30 \\
        $F_{as}$ std.\ dev [$\mu s$] & 53 & 123 & 104 & 25 \\
        Holdover [$\mu s/s$] & -1.399 & 1.658 & -0.032 & -0.371 \\
        \hline
    \end{tabular}
\end{table}

Also, Table \ref{tab:statistics_metrics} reports the Akaike Information Criterion (AIC) normalized by sample size, Bayesian Information Criterion (BIC), and Kolmogorov-Smirnov (KS) statistic computed on the monitored SV streams for one (v)IED per testbed, comparing the EMG distribution against six alternative parametric distributions, i.e., Log-Normal, Gamma, Inverse Gaussian, Weibull, and Generalized Extreme Value (GEV), and a non-parametric Kernel Density Estimator (KDE) with improved Sheather-Jones bandwidth selection as an upper-bound benchmark. Since AIC and BIC are model selection criteria defined for parametric models with a fixed and well-defined parameter count, KDE values are reported for reference only (denoted by *) and are not directly comparable. The results support the choice of EMG across all three testbeds. On the first testbed, EMG achieves the best AIC (9.31), BIC (372), and KS statistic (0.064) among all parametric distributions, with Log-Normal as the closest alternative. On the second testbed, EMG again achieves the best AIC (8.22), BIC (329), and KS statistic (0.167), with GEV as the closest alternative. On the third testbed, GEV marginally outperforms EMG across all three metrics; however, this advantage must be weighed against two important limitations. First, GEV maximum likelihood estimation is numerically unstable, as evidenced by its performance on vIED\textsubscript{1} (KS = 0.601) compared to its good fit on vIED\textsubscript{HSR} (KS = 0.033); this sensitivity to the characteristics of the fitted data makes it unreliable as a general-purpose model across varying network configurations and conditions. Second, GEV parameter estimation has no closed-form solution and requires iterative numerical optimization, which is incompatible with the strict latency and throughput requirements of IEC 61850-compliant IED deployments and would undermine the real-time operation of the proposed intrusion prevention method. Finally, the Gamma distribution never outperforms EMG, and the Inverse Gaussian exhibits inconsistent behavior across testbeds, disqualifying both as robust alternatives. As previously discussed in Section \ref{sec:sv_frames_fas}, beyond statistical fit, the choice of EMG is grounded in physical interpretability, providing a justification that distributions such as Log-Normal or GEV cannot offer. Furthermore, as introduced in Section \ref{sec:emg_distribution}, EMG admits efficient, incremental parameter updates via the closed-form MME. Taken together, EMG is the only distribution that combines good and consistent fit across all three testbeds, physical interpretability, closed-form estimation, and suitability for real-time deployment in IEC 61850-compliant digital substations.

\begin{table}[!t]
\caption{AIC, BIC, and KS test results for EMG and other alternative parametric and non-parametric probability distributions fitted on SV streams on one (v)IED per testbed.}
\label{tab:statistics_metrics}
\centering
    \begin{tabular}{lcccc}
        \hline
        \textbf{Testbed} & \textbf{PDF} & \textbf{AIC} & \textbf{BIC [$\times 10^{3}$]} & \textbf{KS} \\
        \hline
        vIED\textsubscript{1} & \makecell{Gaussian\\Exponential\\EMG\\Log-Normal\\Gamma\\Inverse Gaussian\\Weibull\\GEV\\KDE*} & \makecell{10.29\\10.41\\\textbf{9.31}\\9.35\\9.51\\18.36\\9.84\\11.52\\9.11*} & \makecell{411\\416\\\textbf{372}\\374\\380\\734\\393\\460\\363*} & \makecell{0.186\\0.332\\\textbf{0.064}\\0.065\\0.104\\0.966\\0.146\\0.601\\0.016*} \\
        \hline
        IED\textsubscript{1} & \makecell{Gaussian\\Exponential\\EMG\\Log-Normal\\Gamma\\Inverse Gaussian\\Weibull\\GEV\\KDE*} & \makecell{9.15\\9.19\\\textbf{8.22}\\9.15\\8.49\\9.15\\10.02\\8.24\\7.09*} & \makecell{366\\368\\\textbf{329}\\366\\339\\366\\401\\330\\284*} & \makecell{0.271\\0.370\\\textbf{0.167}\\0.272\\0.200\\0.271\\0.291\\0.187\\0.001*} \\
        \hline
        vIED\textsubscript{HSR} & \makecell{Gaussian\\Exponential\\EMG\\Log-Normal\\Gamma\\Inverse Gaussian\\Weibull\\GEV\\KDE*} & \makecell{12.03\\10.77\\10.61\\12.03\\10.74\\12.03\\13.38\\\textbf{10.55}\\10.46*} & \makecell{24\\22\\22\\24\\22\\24\\27\\\textbf{21}\\21*} & \makecell{0.223\\0.103\\0.082\\0.225\\0.091\\0.225\\0.413\\\textbf{0.033}\\0.016*} \\
        \hline
    \end{tabular}
\end{table}

\begin{figure}[!t]
\centering
\subfloat{%
    \includegraphics[width=0.98\columnwidth]{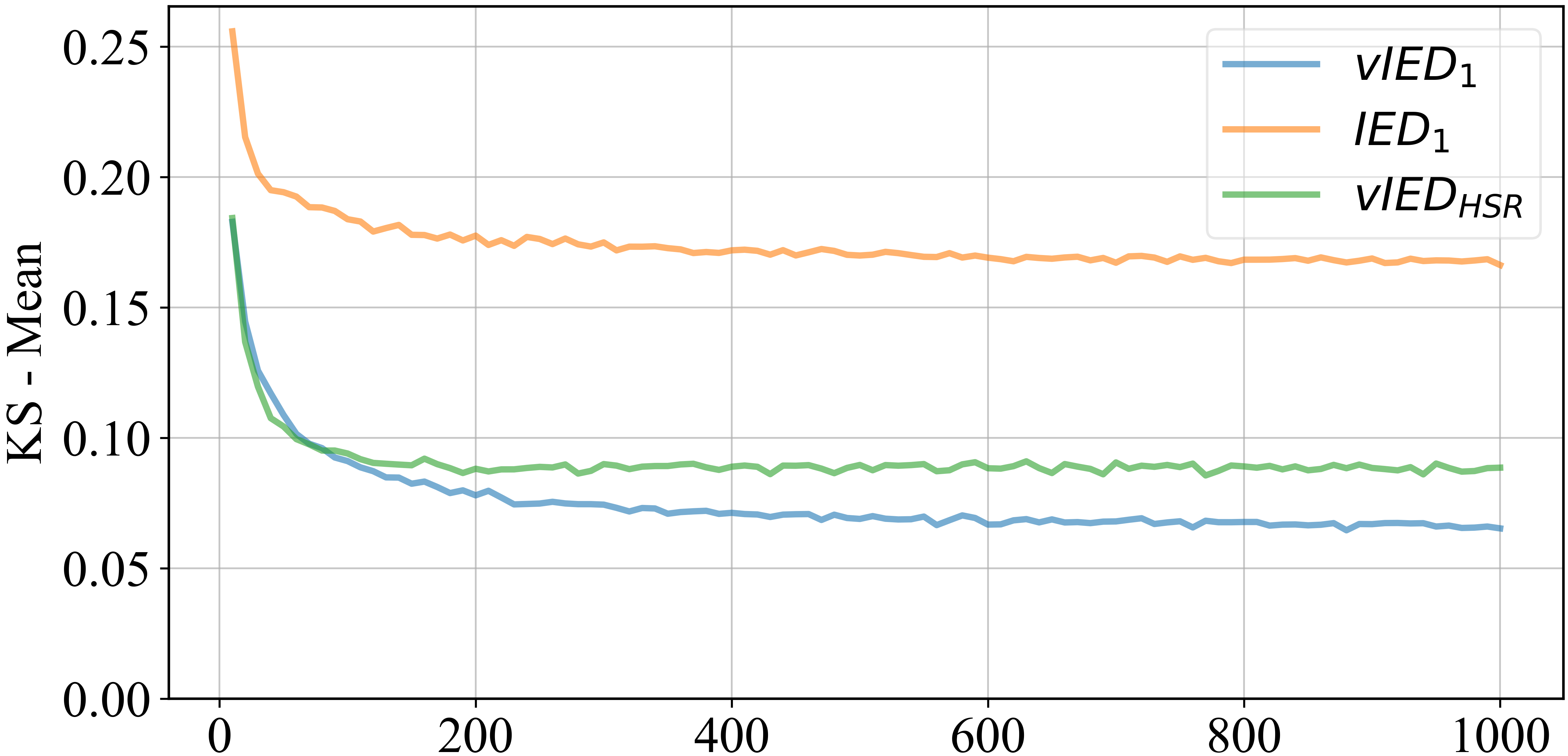}%
    \label{fig_ks_mean}
}\\[1ex]
\subfloat{%
    \includegraphics[width=0.98\columnwidth]{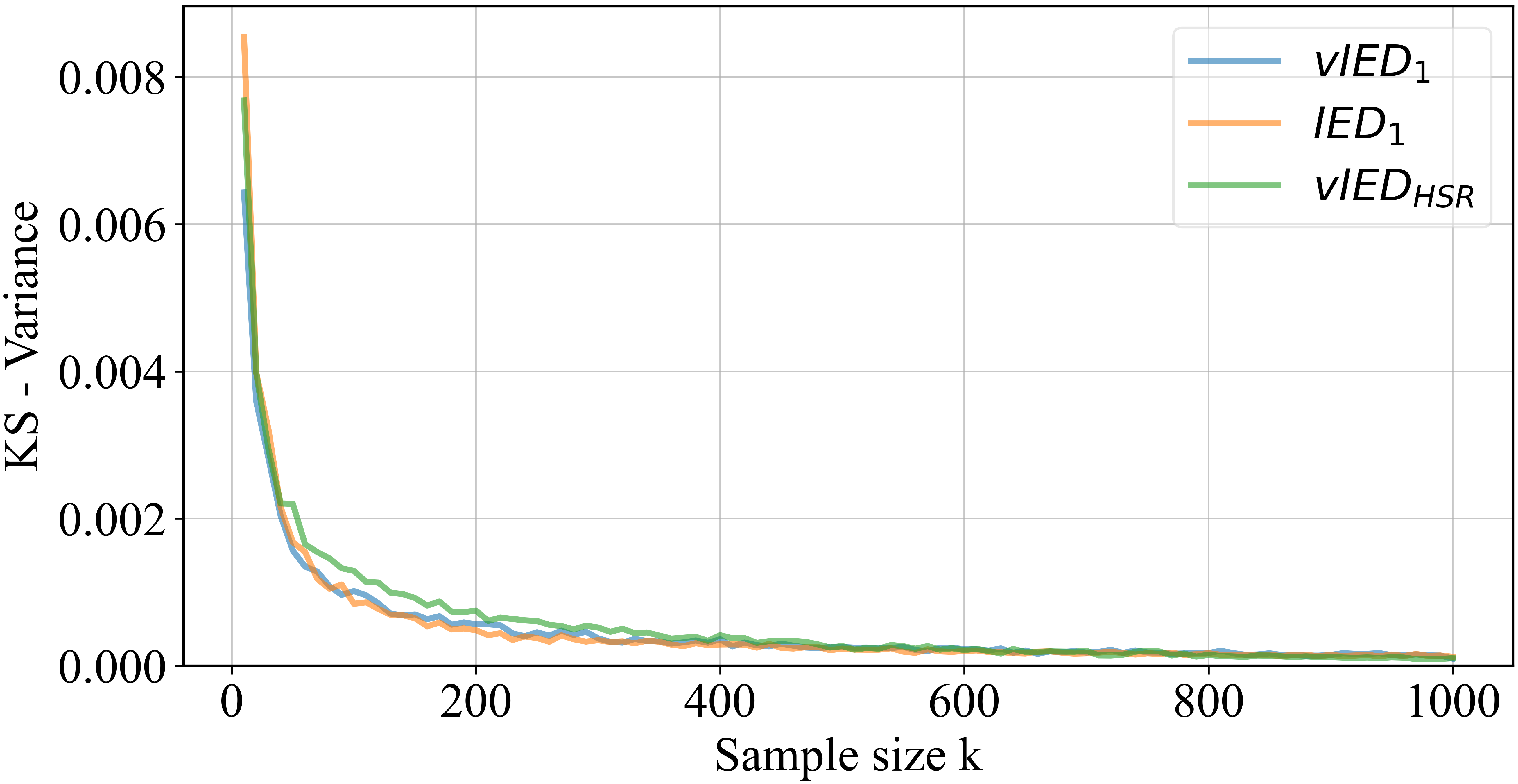}%
    \label{fig_ks_variance}
}
\caption{Sensitivity analysis of the EMG parameter estimation to the sample size $k$. For each value of $k$, the EMG parameters were estimated 500 times by sampling $k$ contiguous frames from one representative SV stream per testbed, and the KS statistic was computed against the empirical distribution.}
\label{fig_ks-score-variance}
\end{figure}

To empirically justify the choice of the value of the parameter $k$ introduced in Section \ref{sec:emg_handler_and_pdf_estimator}, a sensitivity analysis was conducted to assess the stability of the EMG parameter estimation as a function of the sample size. For each value of $k$ in the range [10, 4800], we estimated the EMG parameters 500 times by sampling $k$ contiguous frames from each of the SV streams considered in this work, and computed the mean and variance of the resulting KS statistic against the corresponding empirical distribution. The representative results are reported in Fig. \ref{fig_ks-score-variance} and show two consistent trends across all evaluated datasets. First, both the mean and variance of the KS statistic decrease sharply as $k$ increases from 10 to approximately 400, indicating that small sample sizes yield noisy parameter estimates with high volatility across repetitions. For instance, in the SV stream monitored in the second testbed, the KS mean decreases from 0.256 at $k=10$ to 0.172 at $k=400$, while the KS variance drops from 0.009 to 0.0003 over the same range. Second, beyond $k=400$, both the mean and variance of the KS statistics tend to plateau, with negligible improvements observed up to $k=4000$: the KS mean reaches 0.165 at $k=4000$, a reduction of only 3.96\% relative to $k=400$, but implying an adaptation speed that is 10 times lower. This indicates that $k=400$ lies at the knee of the stability curve, beyond which additional samples yield diminishing returns in estimation accuracy.

\subsection{Intrusion Prevention Evaluation}

To effectively evaluate the performance of the proposed intrusion prevention method, the True Positive Rate (TPR), FPR, and F1-score metrics are considered. For a fair assessment, it is essential to keep into consideration and maximize the TPR while minimizing the FPR. This is because a considerable number of consecutive false negatives can lead the (v)IED to issue unintended control commands to circuit breakers. Conversely, if many legitimate SV frames are discarded because wrongly classified, the protection algorithm is prevented from operating properly and in a timely manner.

Before testing the IPS against SV injection attacks, its performance in terms of false positives during normal operating conditions is evaluated. During normal conditions, the IPS caused FPR of 0.07\%, 0.35\%, 0.32\%, and 0.005\% on vIED\textsubscript{1}, vIED\textsubscript{2}, vIED\textsubscript{3}, and IED\textsubscript{1}, respectively. These false positives were mainly caused by frames being received with an unexpected latency greater than 3 ms. Indeed, as can be noted in Table \ref{tab:sv_streams_info}, higher FPR resulted in (v)IEDs experiencing higher levels of communication jitter, i.e., higher variance of $F_{as}$. These frames exceed IEC 61850 latency limit of 3 ms and would be discarded by protection schemes, aligning with expected IPS behavior. 

Subsequently, malicious SV streams were injected into the OT communication network from a compromised machine connected to the network switch in the first and second testbeds. Due to the attacker’s challenges discussed in Section \ref{sec:attacker_considerations}, perfectly matching the expected $F_{as}$ was not feasible. In fact, after several trials and adjustments, the minimum difference between the frame arrival time shift expected by the SV subscriber ($F_{as}^{leg}$) and frame arrival time shift of the frames injected by the attacker ($F_{as}^{mal}$) was -178 $\mu s$. The difference between $F_{as}^{leg}$ and $F_{as}^{mal}$, which will be referred to as $F_{as}^{(mal-leg)}$, can be seen as the attacker's injection time shift error. As can be appreciated in Table \ref{tab:ips_performance}, even with such small attacker injection time shift errors, the IPS provided FPRs lower than 0.77\% and TPRs higher than 98.5\%.

\begin{table}[!b]
\caption{IPS performances against SV injection attacks performed on the first two testbeds.}
\label{tab:ips_performance}
\centering
    \begin{tabular}{lcccc}
        \hline
        \textbf{} & vIED\textsubscript{1} & vIED\textsubscript{2} & vIED\textsubscript{3} & IED\textsubscript{1} \\
        \hline
        \# Legitimate frames & 43124 & 43124 & 43124 & 50000 \\
        \# Malicious frames & 19201 & 19201 & 19201 & 50000 \\
        $\mathbb{E}[F_{as}^{(mal-leg)}]$ [$\mu s$] & -178 & -223 & -205 & 1001 \\
        FPR [\%] & 0.67 & 0.42 & 0.77 & 0.005 \\
        TPR [\%] & 98.50 & 99.74 & 98.78 & 100 \\
        Precision [\%] & 98.49 & 99.07 & 98.28 & 100 \\
        F1-score [\%] & 98.49 & 99.41 & 98.53 & 100 \\
        \hline
    \end{tabular}
\end{table}

To accurately estimate the IPS performance as a function of $F_{as}^{(mal-leg)}$, a synthetic dataset was generated from the original one. In this synthetic dataset, the original legitimate frames were duplicated, and the $F_{as}$ values of the duplicated frames were varied such that $F_{as}^{(mal-leg)}$ lies within ± 600 $\mu s$. These duplicated frames represent the malicious frames injected by the attacker. This approach allows to accurately identify method's prevention boundaries by analyzing its performance in the relevant search space that reflects attacker capabilities.

\begin{figure}[!t]
\centering
\subfloat{%
    \includegraphics[width=\columnwidth]{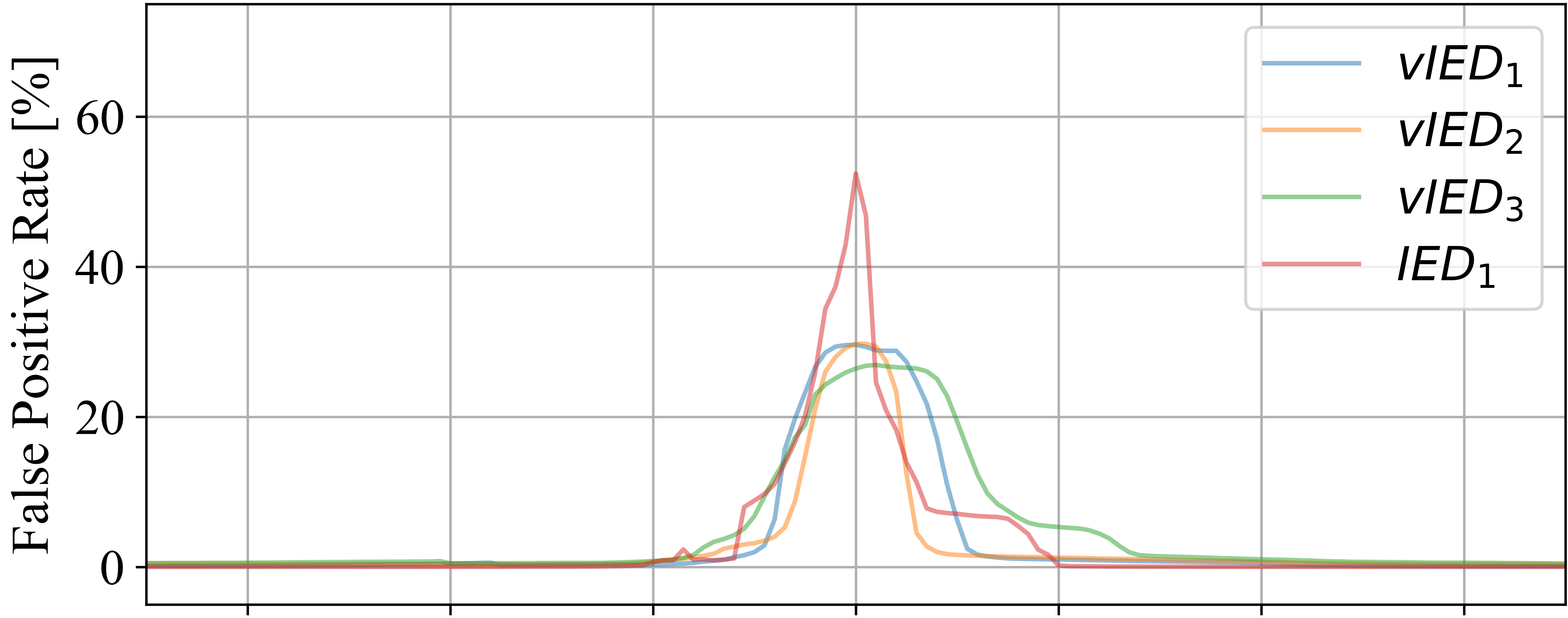}%
    \label{fig_first_case}
}\\[1ex]
\subfloat{%
    \includegraphics[width=\columnwidth]{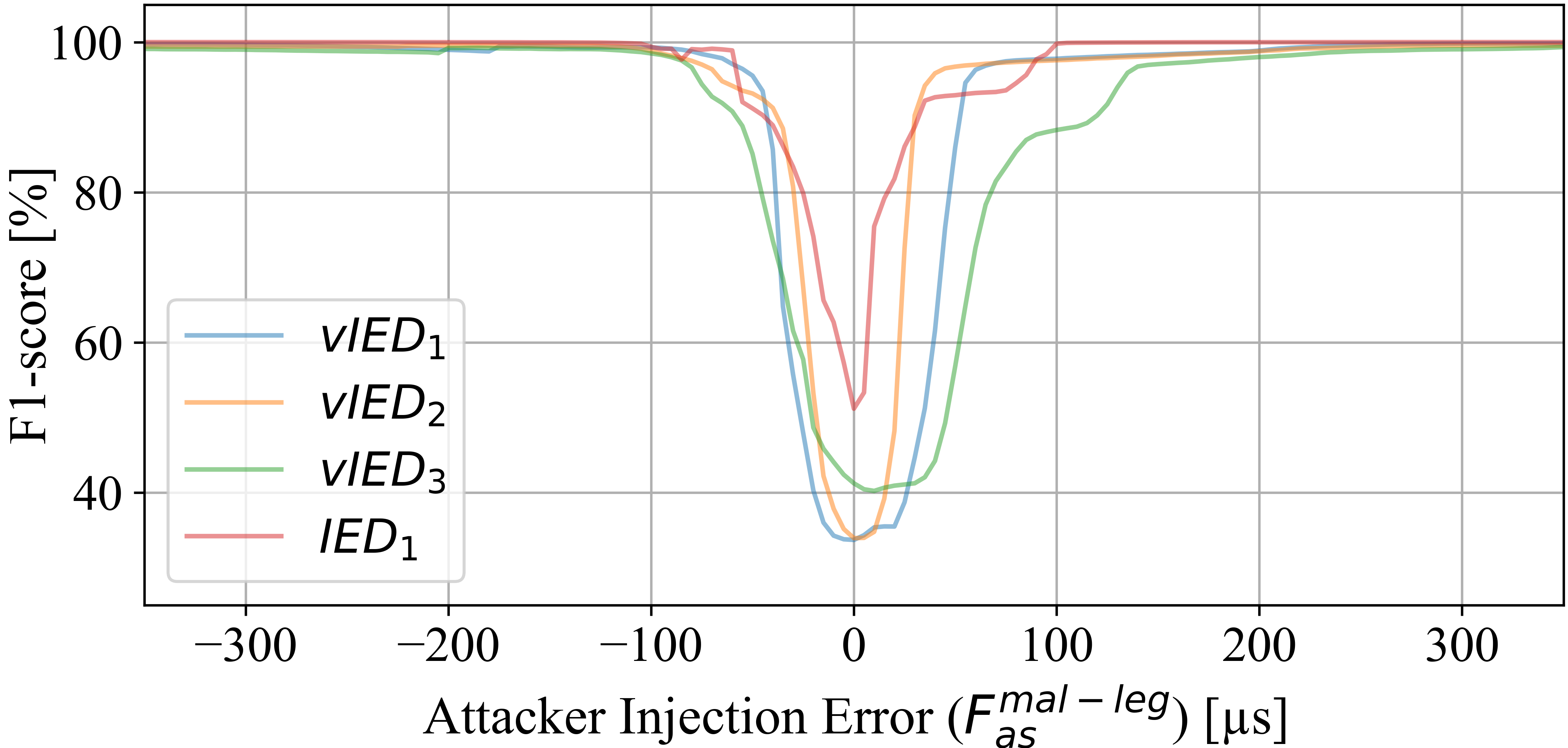}%
    \label{fig_second_case}
}
\caption{FPR and F1-score of the intrusion prevention method as a function of attacker injection error on the four (v)IEDs deployed across the first and second testbeds.}
\label{fig_fpr-f1score}
\end{figure}

As can be observed in Fig. \ref{fig_fpr-f1score}, the IPS provided great prevention capabilities down to an attacker injection error of less than ± 100 $\mu s$, with FPRs lower than 1\% and F1-scores above 95\%. Higher $F_{as}^{(mal-leg)}$ absolute values were not considered because they would lead to F1-scores close to 100\%. As expected, the limitation of the proposed IPS comes into play when the attacker matches the expected legitimate $F_{as}$. In that case, the information advantage is lost, and legitimate and malicious frames have a 50\% probability of being accepted or discarded. Final consideration regards the IPS added latency and throughput performances. Throughout the tests, it was verified that the IPS introduced an additional latency of around 150 $\mu s$ and sustained a throughput exceeding 100,000 frames per second. These performances satisfy the stringent timing requirements specified in IEC 61850-5 and support the highest SV frames publishing rate of 96,000 frames per second, as defined in IEC 61869-9.

\subsection{MitM Detection Evaluation}

Experiments on the first testbed revealed that MitM attacks against a (v)IED produced significant changes in SV frame arrival-time statistics. Specifically, frame interception and tampering increased latency by 40.55 $\mu s$, jitter by 8.48\%, and PDF skewness by 115.79\%, on average. These results demonstrate that SV frame arrival-time statistics provide a reliable indicator for MitM attack detection.

Consequently, the MitM attack detection component was evaluated on four synthetic datasets generated from the network traffic acquisitions from the first two testbeds, enabling the identification of the component's detection boundaries. The attacker, by compromising a forwarding device and performing a MitM attack against the subscriber, could drop the legitimate SV stream and inject a malicious one in real-time. As a result, the SV subscriber received exactly the expected number of SV frames, i.e., $FS$, but the statistical properties of their arrival times were altered due to the MitM attack. This expected behaviour was experimentally verified in the third testbed, where the MitM attacks were practically performed.

\begin{figure}[!b]
\centering
\includegraphics[width=0.88\linewidth]{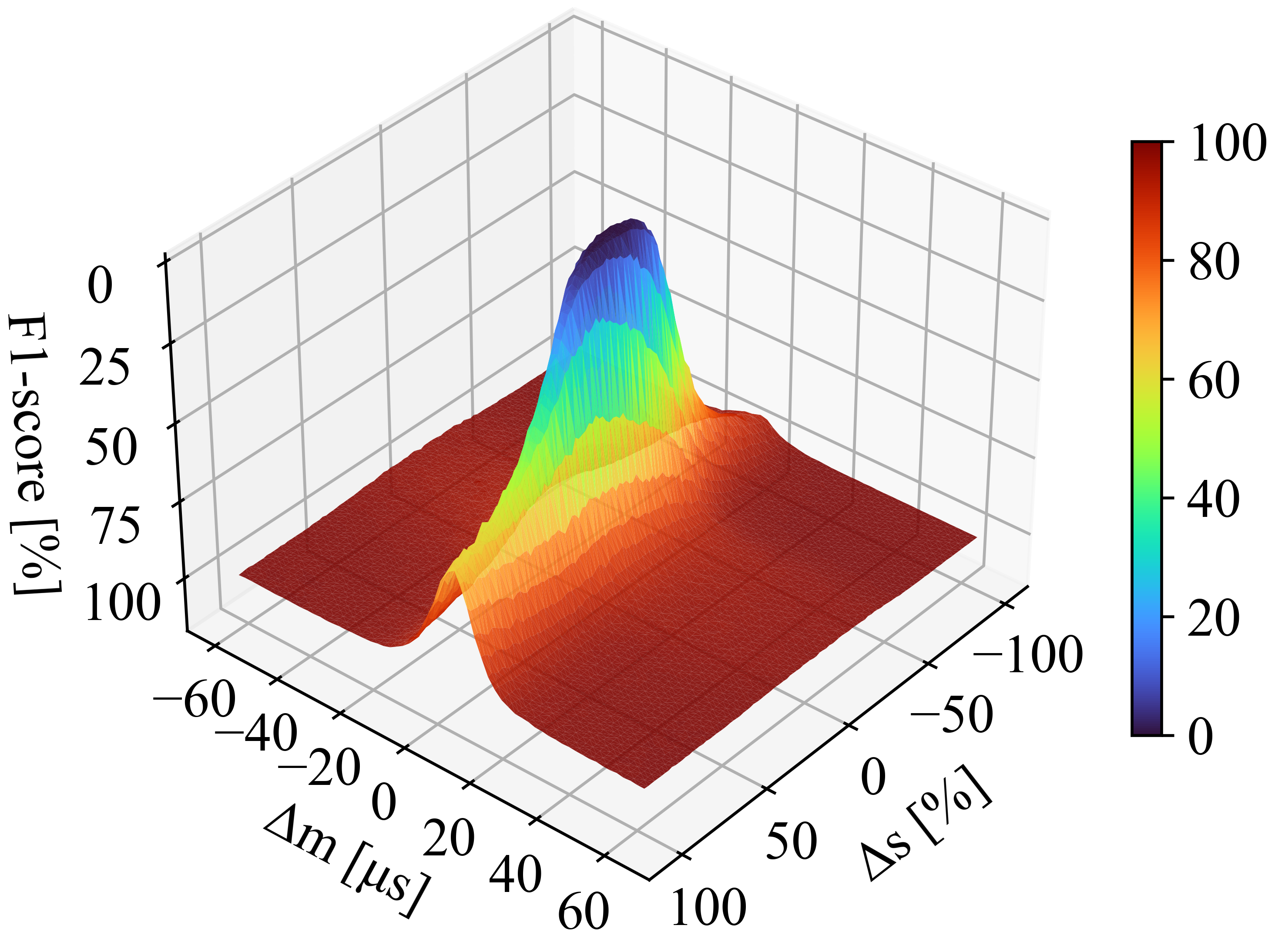}
\caption{F1-score of the MitM detection component in function of the alteration in $F_{as}$ mean and variance due to the cyber attack.}
\label{fig_mitm}
\end{figure}

To generate the synthetic anomalous acquisitions, half of the legitimate data was modified by introducing variations in the mean, standard deviation, and skewness of the measured $F_{as}$. These variations are referred to as $\Delta m$, $\Delta s$, and $\Delta \gamma_1$, with values drawn from uniform distributions with lower and upper bounds of ± 300 $\mu s$, ± 10\%, and ± 5\%, respectively. After selecting a detection threshold providing FPRs lower than 0.15\% and 0.01\% in the first and second testbeds, respectively, the MitM component provided TPRs of 95.15\%, 95.99\%, 93.85\%, and 97.66\%, for vIED\textsubscript{1}, vIED\textsubscript{2}, vIED\textsubscript{3}, and IED\textsubscript{1}, respectively. Still, it must be kept in consideration that in the generated synthetic dataset, some malicious samples were indistinguishable from the legitimate ones, due to random choosing of $\Delta m$, $\Delta s$, and $\Delta \gamma_1$ values close to zero. In Fig. \ref{fig_mitm}, the evaluation results for the second testbed IED with varying $\Delta m$ and $\Delta s$ are depicted. Similar consideration can be drawn for the first testbed vIEDs. As can be appreciated, the MitM detection method provided an F1-score higher than 99.9\% down to an $\Delta m$ variation of 20 $\mu s$. This means that, to remain undetected, an attacker needs to alter the $F_{as}$ measured by less than 20 $\mu s$ on average. Moreover, even if a lower alteration is caused on the expected $F_{as}$ mean, variations in the measured standard deviation or skewness of $F_{as}$ still provide an information advantage to detect the ongoing MitM attack.

\subsection{Attack Source Localization Evaluation}
\label{sec:localization_eval}

\begin{table}[!b]
\caption{Assignment of each scenario's normal and attack data acquisitions to the training and testing datasets.}
\label{tab:scenario_split}
\centering
    \begin{tabular}{lccccc}
        \hline
        \textbf{Scenario ID} & \textbf{1} & \textbf{2} & \textbf{3} & \textbf{4} & \textbf{5} \\
        \hline
        Training set & \makecell{normal\\attack} & -- & \makecell{normal\\attack} & \makecell{normal\\attack} & normal \\
        Testing set     & -- & \makecell{normal\\attack} & -- & --& attack \\
        \hline
    \end{tabular}
\end{table}

This last component was validated on network traffic acquired from the third testbed, considering HSR ring configurations with a varying number of vIEDs, from 5 to 12. To perform MitM attacks, each vIED in the HSR ring was compromised one at a time, its NICs’ HSR bridge was disabled, and a C script was used to monitor, tamper, and forward SV frames in real-time. The network traffic was monitored and acquired with Wireshark at each vIED NIC across multiple experimental scenarios. Each scenario consisted of one case representing normal operating conditions and a number of cases with a MitM attack in place in one vIED, equal to the number of vIEDs in the ring, e.g., for the 8 vIEDs configuration, 8 distinct MitM attack cases were considered. Each scenario was simulated five times. This resulted in 380 acquisitions in total, used for the training and testing of the DL model. As for the validation of the MitM detection component, measured $F_{as}$ mean, standard deviation, and skewness were computed for each vIED, and then delivered to the attack source localization component. To further increase the realism of the experiments, the compromised vIED was allowed to report either normal or abnormal statistical metrics from both NICs; in the following, this process is referred to as ``fare-reporting augmentation.''
The DL model was trained on normal and anomalous cases from a subset of the experimental scenarios and then tested on normal and anomalous cases from different scenarios, as reported in Table \ref{tab:scenario_split}. This setup further demonstrates the generalization capabilities of the proposed DL-based method.

\begin{figure}[!t]
\centering
\includegraphics[width=0.73\linewidth]{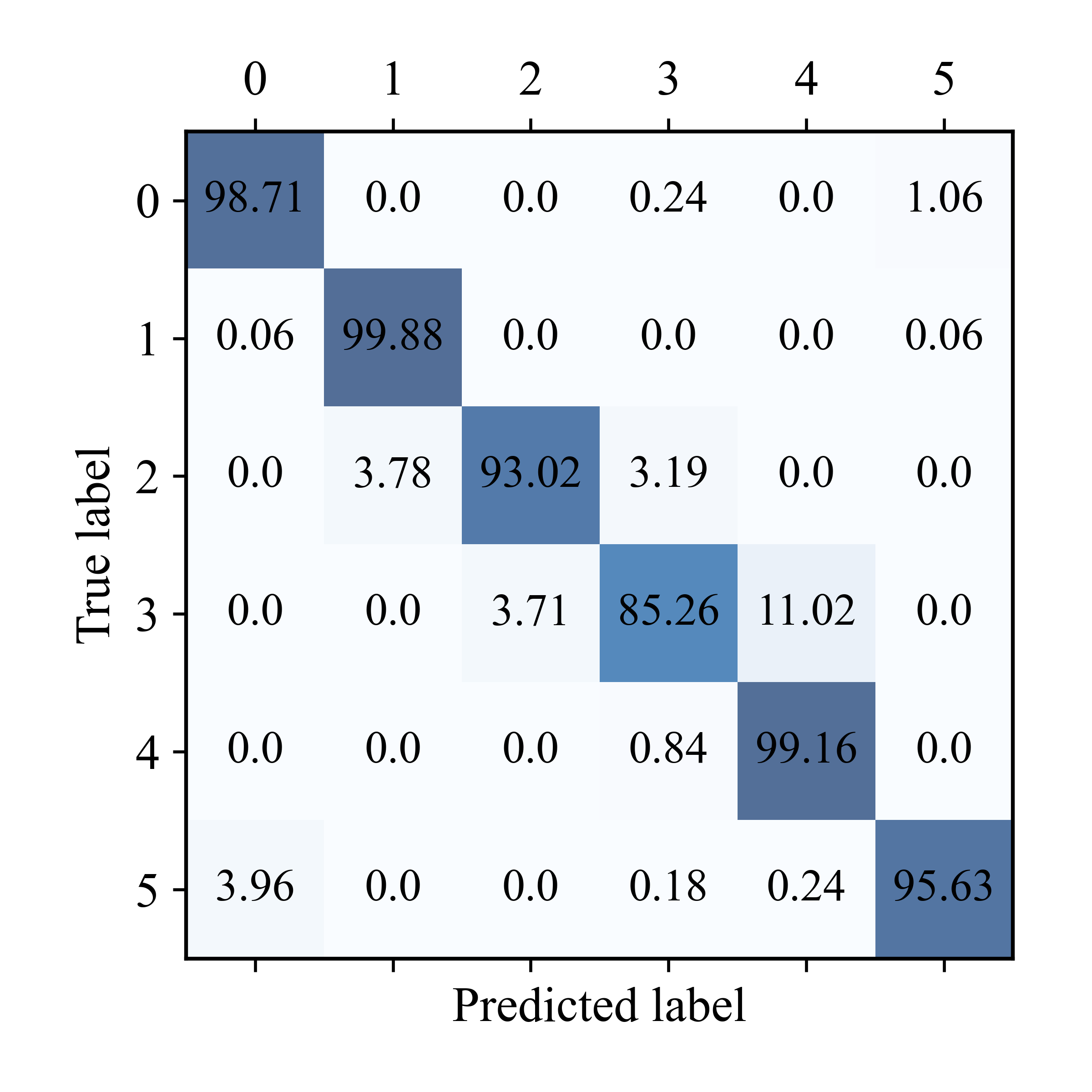}
\caption{Confusion matrix for attack source localization component on the testing set. Label ``0'' corresponds to normal conditions; labels ``1'' to ``5'' indicate which vIED is performing the MitM attack in the third testbed.}
\label{fig_confusion}
\end{figure}

In fact, in a real deployment scenario, only normal operating conditions can usually be acquired from the field and used for training. However, as demonstrated in our experiments, the training set can be expanded with anomalous data acquired from an emulated network. The resulting trained model is then effective in detecting and localizing unseen MitM attacks in the real OT communication network. Table \ref{tab:localization_results} reports the accuracy, precision, recall, and F1-score achieved by the attack source localization method in the configuration deploying 5 vIEDs in the HSR ring, with values above 99\% and 94\% across all metrics in the training and testing datasets, respectively. Furthermore, as can be appreciated in Fig. \ref{fig_confusion}, a FPR of 1.29\% was obtained on the testing dataset. When the attack source was mislocalized, the error was most often confined to devices adjacent to the malicious one, which limits the number of devices requiring further investigation after a cyber security incident.

\begin{table}[!b]
\caption{Attack source localization results for the 5 vIED HSR ring configuration (all values in \%). All methods are evaluated on the same data split, with the fixed fake-reporting augmentation applied to the training and testing data. $^{\dagger}$Original two-stage configuration of~\cite{presekal2023attack}: the GC-LSTM is trained for traffic prediction on normal data and then kept frozen while the classifier is trained on its output. $^{\ddagger}$The GC-LSTM is instead fine-tuned end-to-end together with the classifier.}
\label{tab:localization_results}
\centering
    \begin{tabular}{lcccc}
        \hline
        \textbf{Method} & \textbf{Accuracy} & \textbf{Precision} & \textbf{Recall} & \textbf{F1-score} \\
        \hline
        \multicolumn{5}{l}{\textit{Training set}} \\
        \quad \textbf{Proposed method} & \textbf{99.84} & \textbf{99.83} & \textbf{99.83} & \textbf{99.83} \\
        \quad Rule-based \cite{hong_integrated_2014} & 21.22 & 10.90 & 17.01 & 6.70 \\
        \quad GC-LSTM$^{\dagger}$ \cite{presekal2023attack} & 54.88 & 56.72 & 54.56 & 54.43 \\
        \quad GC-LSTM$^{\ddagger}$ \cite{presekal2023attack} & 88.40 & 89.68 & 88.30 & 88.27 \\
        \hline
        \multicolumn{5}{l}{\textit{Testing set}} \\
        \quad \textbf{Proposed method} & \textbf{94.95} & \textbf{94.88} & \textbf{95.28} & \textbf{94.96} \\
        \quad Rule-based \cite{hong_integrated_2014} & 8.86 & 8.31 & 16.49 & 3.44 \\
        \quad GC-LSTM$^{\dagger}$ \cite{presekal2023attack} & 38.76 & 40.48 & 39.77 & 39.15 \\
        \quad GC-LSTM$^{\ddagger}$ \cite{presekal2023attack} & 70.22 & 72.36 & 71.78 & 69.04 \\
        \hline
    \end{tabular}
\end{table}

To quantitatively compare the proposed method against the state of the art, and given that no prior work targets MitM source localization in HSR-based digital substations under the stealthy conditions considered here, we re-implemented the detection and localization algorithms underlying the most closely related approaches and evaluated them on the 5 vIED configuration. Specifically, we considered the rule-based SV anomaly detection of Hong et al. \cite{hong_integrated_2014}, on which the SDN-based localization frameworks~\cite{girdhar2024sdn, girdhar2025sdn} rely, and the communication-throughput Graph Convolutional-LSTM (GC-LSTM) with graph-based localization of Presekal et al. \cite{presekal2023attack}. Since the SDN infrastructure required by these frameworks is not available in our testbeds, we isolated their detection and localization logic and applied it to the per-interface SV traffic captured across the HSR ring. The rule-based method flags an IED whenever the number of SV frames received per second, the continuity of the \texttt{smpCnt} field, or the source/destination MAC addresses and Ethernet protocol type fields deviate from their expected values. The GC-LSTM method builds a per-interface throughput time series over the ring graph, trains a GC-LSTM to model normal traffic, and classifies each interface as normal or anomalous by means of a convolutional time-series classifier, attributing the attack to the flagged IED. For the latter, we report both the original two-stage configuration of~\cite{presekal2023attack}, in which the GC-LSTM is trained for traffic prediction and then frozen while the classifier is trained on its output, and a stronger non-frozen variant in which the GC-LSTM is fine-tuned end-to-end for classification. The fake-reporting augmentation is applied to each method in its own feature space, i.e., the compromised device disguises the quantities it reports as those of its ring neighbours: the $F_{as}$ statistics for the proposed method and the per-interface throughput for the GC-LSTM. It leaves the rule-based method unaffected, since its decision variables are computed directly from the observed frames and are therefore not altered by falsified reporting.

As reported in Table~\ref{tab:localization_results}, the rule-based method is not able to localize the source of the attack, with an F1-score of 3.44\% on the testing set: because the MitM attacker preserves the SV publishing rate, increments the \texttt{smpCnt} field consistently, and spoofs the legitimate publisher identity, none of the inspected fields deviate and almost all windows are classified as normal. Its accuracy essentially equals the fraction of normal windows in each dataset, i.e., 20.9\% and 8.7\% for the training and testing sets, respectively, which explains the apparent difference between the two rows despite the method having no trainable parameters. The throughput-based GC-LSTM outperforms the rule-based method but remains far below the proposed method. In its original frozen configuration, it reaches only 39.15\% F1-score on the testing set: since the MitM attack preserves the expected throughput, the GC-LSTM, trained to predict traffic volume, normalizes away the small timing perturbations induced by the attack, so its frozen output retains limited discriminative information. Fine-tuning the GC-LSTM end-to-end recovers part of this signal and raises the testing F1-score to 69.04\%, but the method still does not generalize across datasets, and the computed F1-score remains more than 25 percentage points below the one of the proposed method. This confirms that per-interface throughput does not provide a reliable localization signal for stealthy MitM attacks, which preserve traffic volume and perturb only the fine-grained arrival-time statistics. In contrast, the proposed method, which explicitly models the arrival-time fingerprint $F_{as}$ and its spatial-temporal correlations across devices, achieves an F1-score of 94.96\% on the testing set.

\begin{figure}[!t]
\centering
\includegraphics[width=0.99\linewidth]{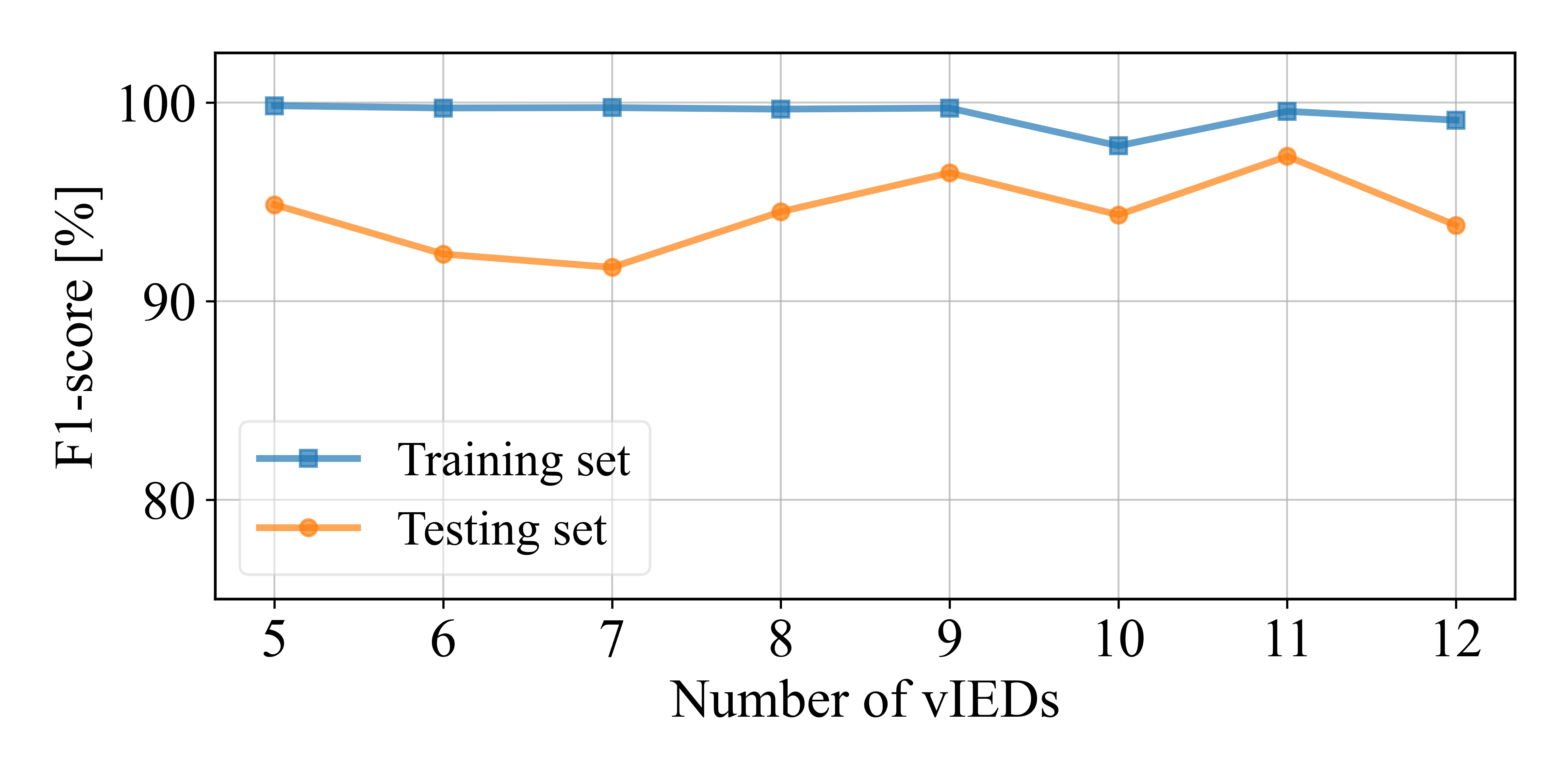}
\caption{F1-score of the attack source localization method as a function of the number of vIEDs deployed in the HSR ring during DL model training and testing.}
\label{fig_f1-score_vIEDs_hsr}
\end{figure}

As can be appreciated in Fig. \ref{fig_f1-score_vIEDs_hsr}, the method also exhibits strong generalization capabilities across HSR-ring sizes. The F1-score computed on the testing set remained consistently above 90\% for ring configurations of up to 12 vIEDs, with localization accuracy across scenarios mainly affected by network operating conditions rather than increased number of deployed vIEDs. Following the network engineering guidelines of IEC 61850-90-4, the HSR ring size in our experiments was capped to 12 vIEDs. These guidelines limit the number of hops between any two communicating IEDs to five for fast trip protection signals. Since HSR forwards frames in both directions around the ring, the worst-case hop count along the shorter path is $\lfloor N/2 \rfloor$, with $N$ the number of IEDs in the ring. A ring of more than 12 IEDs would therefore exceed the five-hop limit and break the 3 ms transfer time required for GOOSE-based protection trip.

\subsection{Final Discussion and Considerations}

Table \ref{tab:protection_capability} summarizes the intrusion prevention capabilities of different methods against the cyber attacks introduced in Section \ref{sec:sv_cybersec}. The methods included in the comparison were selected as those that specifically target intrusion detection for IEC 61850 SV streams in digital substations, rather than general-purpose IDSs or methods designed for other protocols, e.g., GOOSE or PTP, or other OT domains. This selection ensures a focused and meaningful comparison along the dimensions of attack coverage and protection capability that are most relevant to the contribution of this paper. As can be appreciated, the method proposed in this paper addresses the limitations of conventional IDSs by effectively detecting and preventing more advanced attacks.

\begin{table}[!b]
\caption{Existing and proposed solutions' intrusion prevention capability against different types of attacks, where \cmark\ denotes effective prevention, \xmark\ ineffective prevention, and \pmark\ partial prevention.}
\label{tab:protection_capability}
\centering
\begin{tabular}{lcccccc}
\hline
 & \textbf{(1)} & \textbf{(2)} & \textbf{(3)} & \textbf{(4)} & \textbf{(5)} & \textbf{(6)} \\
\hline
Ustun et al. \cite{ustun_artificial_2021}       & \xmark & \cmark & \xmark & \xmark & \pmark & \cmark \\
Mo et al. \cite{mo_sampled_2023}             & \xmark & \cmark & \xmark & \xmark & \pmark & \cmark \\
El Hariri et al. \cite{el_hariri_iec_2019}& \xmark & \cmark & \xmark & \xmark & \pmark & \cmark \\
Hussain et al. \cite{hussain_novel_2023}   & \xmark & \cmark & \pmark & \xmark & \pmark & \cmark \\
Narag et al. \cite{narag2024deep}   & \xmark & \cmark & \pmark & \xmark & \pmark & \cmark \\
Hong et al. \cite{hong_intelligent_2019}         & \pmark & \xmark & \cmark & \cmark & \xmark & \xmark \\
Delhomme et al. \cite{delhomme2024dos} & \pmark & \xmark & \xmark & \xmark & \xmark & \xmark \\
Wannous et al. \cite{wannous_analysis_2019}   & \xmark & \xmark & \cmark & \xmark & \xmark & \xmark \\
Hong et al. \cite{hong_integrated_2014,hong_detection_2014} & \pmark & \xmark & \cmark & \xmark & \xmark & \xmark \\
Eynawi et al. \cite{eynawi_machine_2024}     & \xmark & \xmark & \cmark & \xmark & \xmark & \xmark \\
Manzoor et al. \cite{manzoor_zero-day_2024}   & \xmark & \xmark & \cmark & \cmark & \xmark & \xmark \\
\hline
\textbf{Our Solution}               & \textbf{\cmark} & \textbf{\cmark} & \textbf{\cmark} & \textbf{\cmark} & \textbf{\cmark} & \textbf{\cmark} \\
\hline
\end{tabular}
\end{table}

During flooding attacks, conventional IDSs can detect SV frames reception rate anomalies and raise alerts, but they lack mechanisms to forward only legitimate frames to IEDs; therefore, the targeted device still receives the malicious frames and must process them, leading to resource exhaustion regardless of whether an alert has been raised. The proposed method addresses this gap by discerning legitimate and malicious frames at the per-frame level and ensuring that at most $FS$ frames per second are received by the protected IED’s protection logic, thus drastically reducing the targeted device’s computational load and effectively mitigating, rather than merely detecting, the flooding attack.

For replay attack injecting previously sniffed SV frames, the arrival PDF and replay protection module detects and discards duplicated \texttt{smpCnt} values. Although alternative IDSs monitoring \texttt{smpCnt} can identify such attacks, they cannot differentiate legitimate frames from malicious duplicates. Further, IDSs relying on current and voltage measurements monitoring remain ineffective because injected values may remain consistent with normal power system operation.

As shown in the quantitative analysis of the experimental scenarios, spoofing attacks are prevented by discarding the SV frames with the lowest legitimacy likelihood. The inherent difficulty of accurately and consistently estimating $F_{as}$ prevents attackers from successfully mimicking legitimate traffic, thereby enabling effective detection and prevention of spoofing attacks. In contrast, other IDSs that rely on frame-rate monitoring can detect spoofing when both legitimate and malicious SV streams are received, but they cannot distinguish which frames are legitimate.

High \texttt{smpCnt} attacks, which trick replay protection by injecting frames with artificially elevated \texttt{smpCnt} field values, are similarly thwarted because the frame arrival time constraints remain unsatisfied, causing malicious frames to be discarded.

During MitM attacks, where legitimate frames are dropped and malicious ones are injected at exactly $FS$ frames per second, the arrival PDF and replay protection modules may initially accept malicious frames. However, the MitM detection module identifies the attack because matching the statistical properties, i.e., mean, variance, and skewness, of the legitimate SV stream presents substantial difficulty, particularly under realistic operating conditions. Prior IDS approaches cannot detect such attacks when the frame rate, \texttt{smpCnt} field, SV stream identifier, and inter-frame timing are all properly emulated.

Finally, when SV frame authentication is enabled, but the pre-shared authentication key has been compromised, the proposed method's arrival PDF and replay protection component detects and prevents attacks failing to satisfy the expected frame arrival time shift constraint, ensuring similar performances as during spoofing attack scenarios.

\section{Conclusions}
This paper presents a novel integrated method based on hybrid statistical-deep learning for real-time detection, prevention, and source localization of IEC 61850 SV-based cyber attacks in digital substations. It has been shown that the statistical modeling of SV frames arrival time combined with DL enables the distinction between legitimate and malicious SV frames, early detection of MitM attacks on SV streams, and localization of malicious IEDs within the digital substation communication network. Experimental validation demonstrates promising intrusion detection, prevention, and localization performances while ensuring no interference to protection schemes in the absence of cyber attacks. The method satisfies IEC 61850-5 and IEC 61869-9 strict latency requirements and required throughput, respectively, and is robust to communication network latency, jitter, and time synchronization deviations. Thus, it provides a viable solution to enhance cyber security and resilience in digital substations. Although validated on IEC 61850-compliant testbeds, future work will focus on field deployment in operational digital substations and defenses against advanced attacker strategies designed to evade frame acceptance filtering.

\bibliographystyle{IEEEtran}
\bibliography{bibliography}  

\vfill

\end{document}